\documentclass[aip,reprint,amsmath,amssymb,superscriptaddress,onecolumn,nofootinbib]{revtex4-1}
\pdfoutput=1
\usepackage{graphicx} 
\usepackage{bm} 
\usepackage{dcolumn}
\usepackage{subfigure}
\usepackage{sidecap}
\usepackage{lettrine}
\usepackage{url}
\usepackage[misc]{ifsym}
\usepackage[ 
colorlinks,
linkcolor=blue,
anchorcolor=blue,
citecolor=blue,
urlcolor=blue]
{hyperref}
\usepackage{soul}
\setstcolor{red}
\setulcolor{blue}
\begin{document}

\title{Laboratory observation of plasmoid-dominated magnetic reconnection in hybrid collisional-collisionless regime \smallskip}

\author{Z.~H.~Zhao$^{\dagger}$}
\affiliation{Center for Applied Physics and Technology, HEDPS, and SKLNPT, School of Physics, Peking University, Beijing 100871, China}
\author{H.~H.~An$^{\dagger}$}
\affiliation{Shanghai Institute of Laser Plasma, CAEP, Shanghai 201800, China}
\author{Y.~Xie}
\affiliation{Center for Applied Physics and Technology, HEDPS, and SKLNPT, School of Physics, Peking University, Beijing 100871, China}
\author{Z.~Lei}
\affiliation{Center for Applied Physics and Technology, HEDPS, and SKLNPT, School of Physics, Peking University, Beijing 100871, China}
\author{W.~P.~Yao}
\affiliation{Center for Applied Physics and Technology, HEDPS, and SKLNPT, School of Physics, Peking University, Beijing 100871, China}
\author{W.~Q.~Yuan}
\affiliation{Center for Applied Physics and Technology, HEDPS, and SKLNPT, School of Physics, Peking University, Beijing 100871, China}
\author{J.~Xiong}
\affiliation{Shanghai Institute of Laser Plasma, CAEP, Shanghai 201800, China}
\author{C.~Wang}
\affiliation{Shanghai Institute of Laser Plasma, CAEP, Shanghai 201800, China}
\author{J.~J.~Ye}
\affiliation{Shanghai Institute of Laser Plasma, CAEP, Shanghai 201800, China}
\author{Z.~Y.~Xie}
\affiliation{Shanghai Institute of Laser Plasma, CAEP, Shanghai 201800, China}
\author{Z.~H.~Fang}
\affiliation{Shanghai Institute of Laser Plasma, CAEP, Shanghai 201800, China}
\author{A.~L.~Lei}
\affiliation{Shanghai Institute of Laser Plasma, CAEP, Shanghai 201800, China}
\author{W.~B.~Pei}
\affiliation{Shanghai Institute of Laser Plasma, CAEP, Shanghai 201800, China}
\author{X.~T.~He}
\affiliation{Institute of Applied Physics and Computational Mathematics, Beijing 100094, China}
\author{W.~M.~Zhou$^{\textrm{\Letter}}$}
\affiliation{Science and Technology on Plasma Physics Laboratory, Research Center of Laser Fusion, China Academy of Engineering Physics (CAEP), Mianyang 621900, China}
\author{W.~Wang$^{\textrm{\Letter}}$}
\affiliation{Shanghai Institute of Laser Plasma, CAEP, Shanghai 201800, China}
\author{S.~P.~Zhu$^{\textrm{\Letter}}$}
\affiliation{Institute of Applied Physics and Computational Mathematics, Beijing 100094, China}
\author{B. Qiao$^{\textrm{\Letter}}$}
\affiliation{Center for Applied Physics and Technology, HEDPS, and SKLNPT, School of Physics, Peking University, Beijing 100871, China}
\renewcommand{\thefootnote}{}
\footnotetext{$\dagger$ \  These authors have contributed to this work equally.}
\footnotetext{$\textrm{\Letter}$ \  Corresponding authors: \url{bqiao@pku.edu.cn},  \url{zhu\_shaoping@iapcm.ac.cn}, \url{wei_wang@fudan.edu.cn}, \url{zhouwm@caep.cn}.}

\date{\today}

\begin{abstract}
\bf

Magnetic reconnection, breaking and reorganization of magnetic field topology, is a fundamental process for rapid release of magnetic energy into plasma particles that occurs pervasively throughout the universe. In most natural circumstances, the plasma properties on either side of the reconnection layer are asymmetric, in particular for the collision rates that are associated with a combination of density and temperature and critically determine the reconnection mechanism. To date, all laboratory experiments on magnetic reconnections have been limited to purely collisional or collisionless regimes. Here, we report a well-designed experimental investigation on asymmetric magnetic reconnections in a novel hybrid collisional-collisionless regime by interactions between laser-ablated Cu and CH plasmas. We show that the growth rate of the tearing instability in such a hybrid regime is still extremely large, resulting in rapid formation of multiple plasmoids, lower than that in the purely collisionless regime but much higher than the collisional case. In addition, we, for the first time, directly observe the topology evolutions of the whole process of plasmoid-dominated magnetic reconnections by using highly-resolved proton radiography.

\end{abstract}

\maketitle

Magnetic reconnection is a physical process occurring nearly anywhere there's plasma, in which the magnetic topology is rearranged and magnetic energy is converted to kinetic energy, thermal energy, and particle acceleration \cite{Zweibel2009, Yamada2010}. As plasma makes up the stars and ninety-nine percent of the visible universe, magnetic reconnection is ubiquitous and plays a key role in many energetic events throughout the whole universe such as solar flares \cite{Schrijver1998}, gamma-ray-burst (GRB)\cite{Piran2004} and so on. However, due to the vastly different plasma properties and spatiotemporal scales, it is hard to give a universal model for describing the mechanism of magnetic reconnection.

A key element of magnetic reconnection is the reconnection electric field $\bf E_{rec}$, which plays a pivotal role in both energy conversion and production of energetic particles by doing work on particles. The normalized reconnection electric field is also employed to represent the reconnection rate. Theoretical and simulation studies have shown that $\bf E_{rec}$ depends heavily on the plasma properties\cite{Biskap1997}, in particular, the collision rate as a combination of plasma density and temperature. In the strongly collisional regime, such as magnetic reconnections in solar/stellar photosphere and chromosphere\cite{Shibata2007}, $\bf E_{rec}$ is induced by the plasma resistivity as described by the Sweet-Parker model \cite{Sweet1958, Parker1963}, which, however, leads to a rather slow reconnection with low reconnection rate. By contrast, in the weakly collisional or collisionless regime, such as those occurring in magnetopause\cite{Cassak2016} and magnetotail\cite{Petrukovich2016}, $\bf E_{rec}$ is contributed mainly by the off-diagonal (nongyrotropic) component of the electron pressure tensor \cite{Hesse2001, Hesse2011}, resulting in a fast reconnection that is independent of the plasma collision rate.
 
\begin{figure*}
    \includegraphics[width= 14 cm]{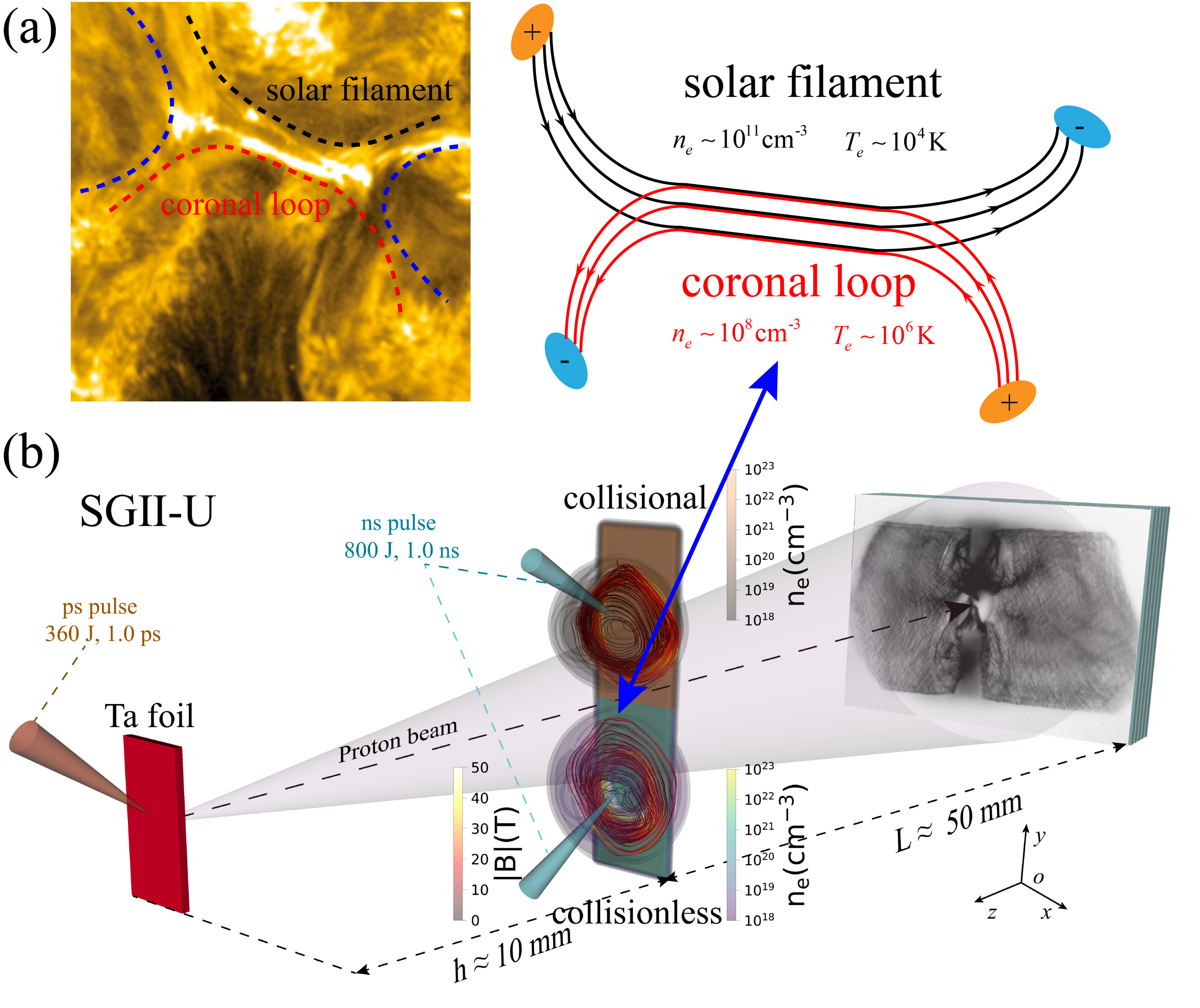}
    \caption{{\bf Hybrid collisional-collisionless magnetic reconnection configurations and experimental set-up.} (a) Magnetic reconnection in the hybrid regime between the collisional solar filaments and the nearby collisionless coronal loops. The original observation data is given by the Solar Dynamics Observatory (SDO), and the picture, observed by Atmospheric Imaging Assembly (AIA) at 171 $\rm \mathring{A}$ ($\sim 0.9$ MK), is re-edited from Li \emph{et al.} \cite{Li2016}. (b) Setup for the experiment on magnetic reconnection in the hybrid collisional-collisionless regime, which is achieved by irradiation of Cu (yellow) and CH ($\rm C_1H_1$, green) foil targets respectively with long ns laser pulses synchronously. Proton radiography is set up along the face-on ($-z$-axis) direction for probing the magnetic field topology changes during the reconnection process, where the high-quality proton beam is generated from a tantalum foil target driven by the relativistic ps laser pulse.}
    \label{fig:1}
\end{figure*}

However, for most natural circumstances in the universe, a large class of magnetic reconnection lies in a hybrid regime where the reconnection plasmas have asymmetric collision rates with one side in a strongly-collisional state and the other in a weakly-collisional or collisionless state. One of the most typical scenarios for such hybrid collisional-collisionless magnetic reconnection occurs in the solar/stellar atmosphere\cite{Li2016, Yang2019}, see Figure \ref{fig:1}(a), when the cool (electron temperature $T_e \sim 10^4$ K), dense (plasma density $n_e \sim 10^{11}$ cm$^{-3}$) filaments erupting from the chromosphere collide with the hot ($T_e \sim 10^6$ K), tenuous ($n_e \sim 10^8$ cm$^{-3}$) loops in the corona. As we know, the collisional mean free path $\lambda_{\rm ei} \propto T_e^2 n_e^{-1}$, therefore, the plasma in the filaments is highly collisional while that in the coronal loops is collisionless. Such hybrid magnetic reconnection may also occur when the magnetic fields in the dense accretion disk collide with those in the tenuous interstellar medium.

So far, most laboratory experiments \cite{Nilson2006, Li2007, Zhong2010, Fiksel2014} of magnetic reconnections are focused on the purely collisional and/or collisionless regimes. The reconnection experiment \cite{Rosenberg2015a, Rosenberg2015b} by laser-driven colliding plasmas with asymmetric flow velocities arising from the delay between laser drives are carried out, where, however, almost no impacts on the reconnection dynamics have been observed because the reconnection plasmas are both still purely collisionless. In addition, to date, the evolutions of magnetic topologies for the whole process of the plasmoid-dominated magnetic reconnections\cite{Shibata2001, Shibata2016} including the growth of tearing instabilities\cite{Uzdensky2016} and the formation of multiple plasmoids\cite{Samtaney2009, Uzdensky2010} have never been directly observed, where only indirect measurements through the interferometry \cite{Dong2012, Hare2017} have been given.

Here, we report result of the first experiment on asymmetric magnetic reconnections in the novel hybrid collisional-collisionless regime by colliding of laser-ablated high-Z Cu and low-Z CH plasmas. It shows that the growth rate of the tearing instability in such a hybrid regime is still extremely large, resulting in rapid formation of multiple plasmoids, lower than that in the purely collisionless regime but much higher than the collisional case (where only single X-point forms). Using the temporally and spatially highly-resolved proton radiography, we provide, for the first time, the direct measurement of magnetic topology evolutions for the whole process of such plasmoid-dominated magnetic reconnection, so that the specific reconnection dynamics including growth of the tearing instability and formation of multiple plasmoids are well characterized. The experimental results are well reproduced and explained by self-consistently combining the radiation-magnetohydrodynamic (RMHD) and particle-in-cell (PIC) simulations as well as the proton radiography iterative inversion algorithm. Physically, in the hybrid magnetic reconnection, the reconnection electric field show much distinct feature, which is large and grows fast at the collisionless plasma side induced by the non-gyrotropic component of the electron pressure tensor, whereas smaller and more slowly at the collisional side induced by only the resistivity. 

\section*{Results} 

\subsection*{Experimental setup}
The hybrid magnetic reconnection experiments are carried out on the ShenGuang II Upgrade (SG-II-U) laser facility that has 8 ns pulses and 1 relativistic ps pulse. Figure \ref{fig:1}(b) shows a diagram of the experimental setup, where the hybrid collisional-collisionless reconnection is achieved by irradiation of Cu and CH foil targets respectively with long ns laser pulses synchronously. The ablated low-Z CH plasma is in the collisionless state, while the high-Z Cu plasma is in the collisional state, due to their different ionization charge states. These two plasmas expand and collide with each other, forming the magnetic reconnection in the hybrid regime, because both of them advect the self-generated Biermann magnetic fields together, as also shown in \ref{fig:1}(b) (see the density maps and field lines). In order to balance the aspect ratio of the current sheet $ L / \delta $ and the relative velocity of the plasma bubbles, the distance between the two focal spots is appropriately set to be 1.8 mm. The temporally and spatially highly-resolved proton radiography is set up along the face-on ($-z$-axis) direction for probing the magnetic field topology changes during the reconnection process, where the high-quality proton beam is generated by target normal sheath acceleration from a tantalum foil target driven by the relativistic ps laser pulse. The static radiography images in Supplementary Fig. \textcolor{blue}{\bf S1} clearly show that the proton beam is of high quality with almost uniform distributions, where the cut-off energy exceeds 30 MeV. For protons with energy of typically 13.5 MeV, we estimate the time for them to pass through the reconnection region is about 20 ps, far less than the characteristic plasma evolution time (generally $\sim$ ns), so that the transient radiography can be guaranteed. 

\begin{figure}
    \includegraphics[width= 8.4 cm]{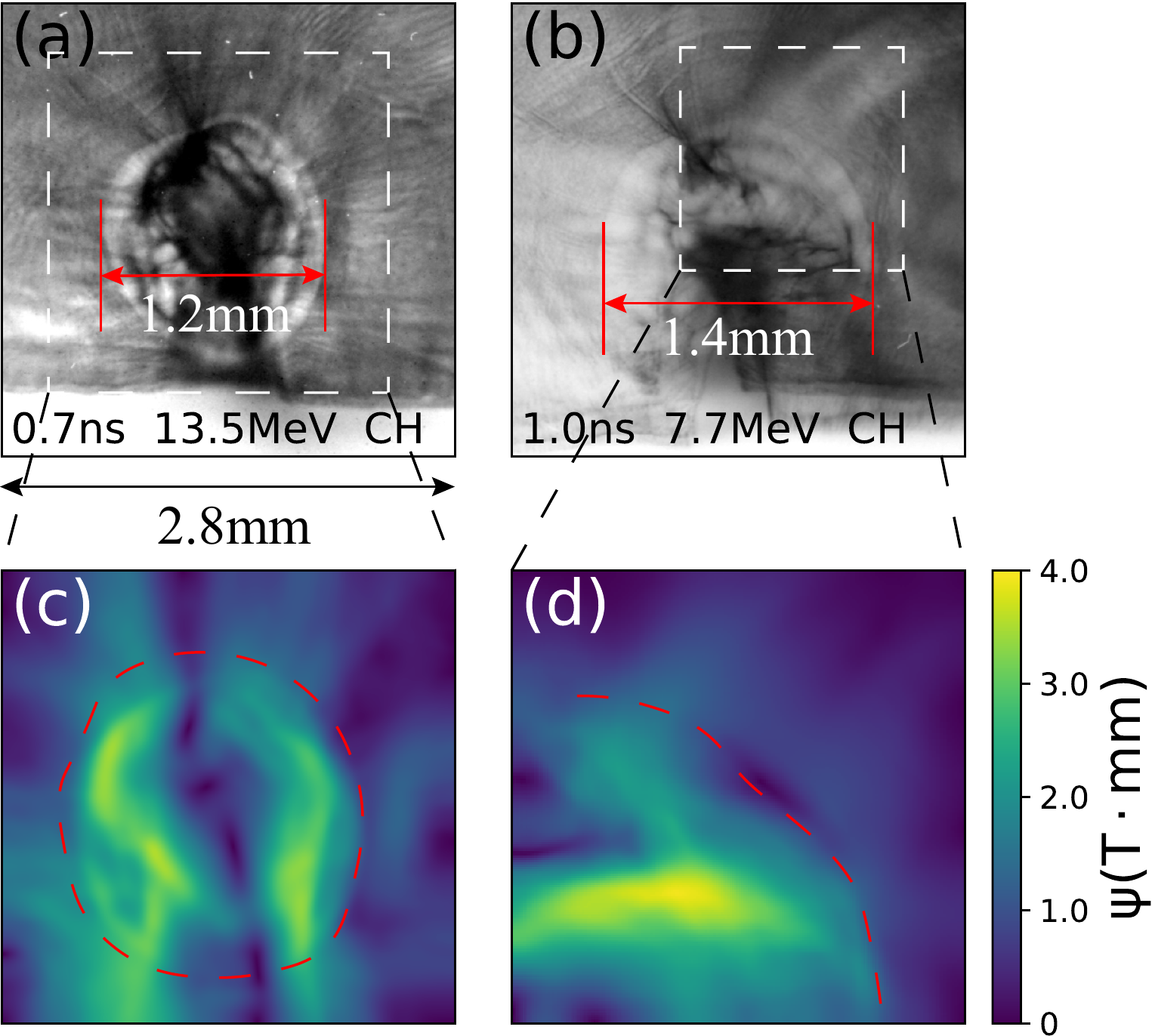}
    \caption{ {\bf Topologies of self-generated Biermann magnetic fields in laser-driven expanding CH plasma bubbles.} (a) and (b) are respectively the face-on proton radiography images at $ t = 0.7 $ ns and $1.0 $ ns for CH plasma expansion, where the higher the grayscale represents the higher the proton doses. (c) and (d) are the strength of the reconstructed path-integrated magnetic fields $\psi$ (in units of $\rm T \cdot mm$), corresponding to the dashed-box region in (a) and (b) respectively.}
    \label{fig:2}
\end{figure}  

\subsection*{Features of self-generated Biermann magnetic fields}
To have an intuitive understanding of the self-generated Biermann magnetic fields in laser-driven expanding plasmas, we firstly take proton radiography for a single CH plasma bubble. The radiography images are shown in Figs. \ref{fig:2}(a) and \ref{fig:2}(b) at time $t=0.7$ and $1.0$ ns, respectively, which correspond to the radiography protons of $13.5$ and $7.7$ MeV. We see that the protons are deflected into the inside of the plasma bubble and a clear low-dose ring structure is formed on the periphery of the plasma bubble, which can be regarded as the direct evidence for generation of a toroidal, clockwise Biermann magnetic field. Note that, if it is a radial electric field, the protons should be deflected outside the plasma bubble \cite{Petrasso2009}. Further, comparing \ref{fig:2}(a) and \ref{fig:2}(b), we can also estimate the plasma expansion velocity parallel to the target surface is about 700 km/s\cite{Gao2015}, as about 2.0 $c_s$, where $c_s \equiv (\gamma \bar{Z} T_e / m_i)^{1/2} $ is the ion sound speed. Similar radiography images of laser-ablated Cu plasmas are also obtained, by which we estimate that its expansion velocity is on the same order due to the similar charge-to-mass ratio $\bar{Z} / A$. These self-generated Biermann magnetic field topologies are also verified by our three-dimensional (3D) synchronous proton radiography experiment \cite{Zhao2021}.

\begin{figure*}
    \includegraphics[width= 16 cm]{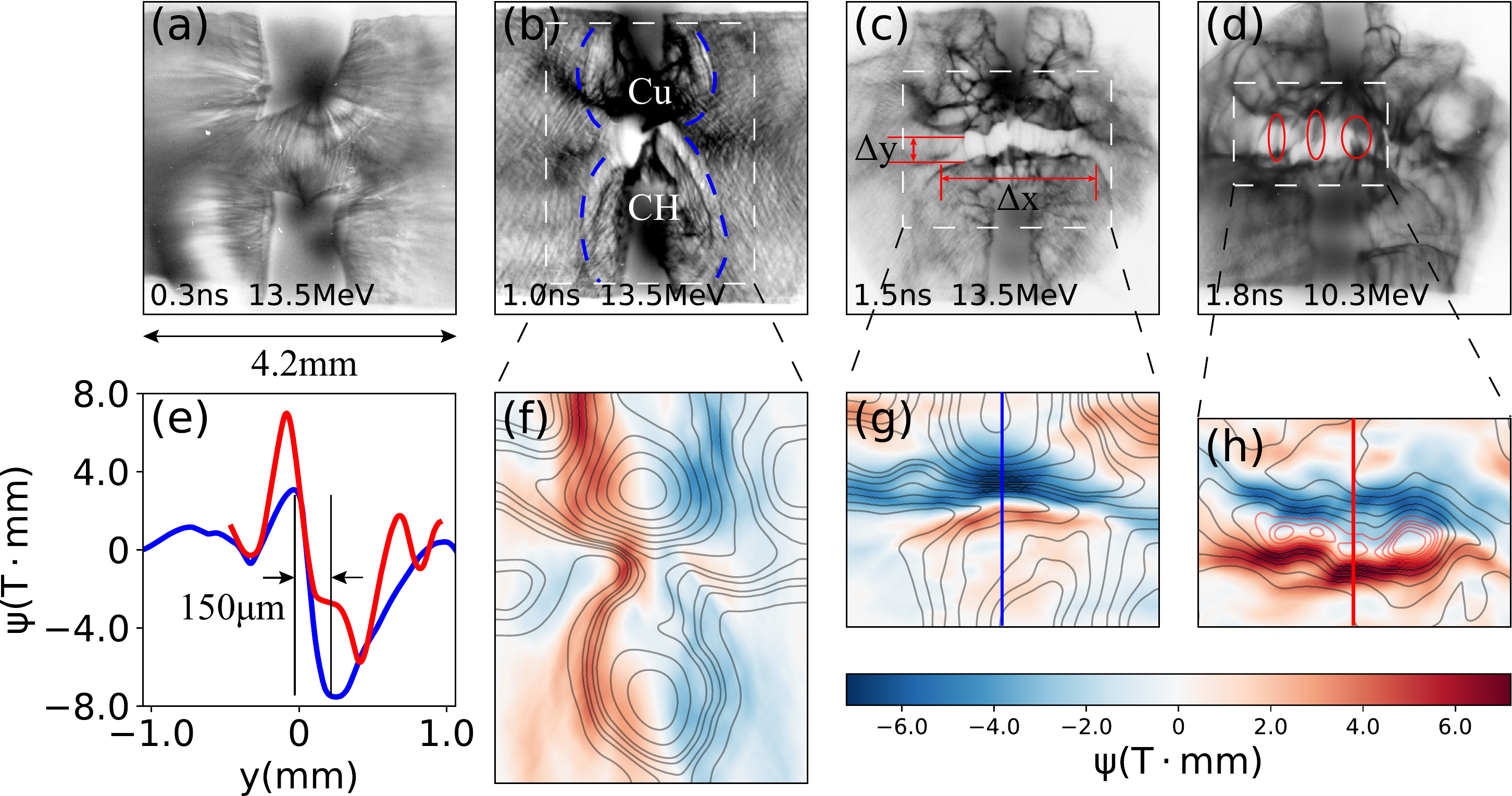}
    \caption{{\bf Experimental results of hybrid collisional-collisionless magnetic reconnection.} (a)-(d) are the proton radiography results of magnetic field topologies at time $t=0.3$, $1.0$, $1.5$ and $1.8$ ns respectively, where the upper half is for collisional Cu plasma and the lower is for collisionless CH. The higher the grayscale represents the higher the proton doses, and all images are rescaled to the realistic length through dividing by the radiography geometric magnification $M \sim 6.0$, see Methods. (f)-(h) are the reconstructed path-integrated magnetic fields $\psi$, corresponding to the dashed-box region in (b)-(d) respectively, where the color map in (f) represents $B_y$ component and those in (g) and (h) represent $B_x$ components. The contour lines in (f)-(h) represent the corresponding vector potential $A_z$. (e) shows the profiles of $\psi$ in (g) (blue line) and (h) (red line) along y-axis.}
    \label{fig:3}
\end{figure*}

To extract more quantitative information from the radiographys, we use the inverse field-reconstruction code ``PROBLEM'' \cite{Bott2017} to recover the path-integrated magnetic fields $\psi$, see Methods. Reconstructed $\psi$ for the dashed-box region in Figs. \ref{fig:2}(a) and \ref{fig:2}(b) are shown in \ref{fig:2}(c) and \ref{fig:2}(d) respectively, which further confirms the existences of toroidal magnetic fields in expanding plasmas (see red dashed curves). The value of $\psi$ is about 4.0 $\rm T \cdot mm$, from which we can estimate the average field strength about $>10$ T. This is in good consistence with the estimation under paraxial approximation, where the path-integrated magnetic field $\psi$ is estimated  \cite{Li2007, Rosenberg2015a, Rosenberg2015b} as 
  \begin{equation}
      \psi ({\rm T \cdot mm }) \equiv \int_0^d \textbf{B} dz \approx 0.1445 \frac{w({\rm \mu m}) \sqrt{E_p ({\rm MeV})}}{L({\rm mm})},
  \end{equation}
where $w$ and $L$ are respectively the width of the low-dose ring and the distance from the plasma to the RCF stack, and $E_p$ is the proton energy. Assuming $w \sim 500 \mu$m, $E_p \sim 13.5 $ MeV and $L \sim 50 $ mm for CH plasma in \ref{fig:2}(a), the magnetic field is roughly estimated as $\psi \approx 5.3 $ $\rm T \cdot mm$, and the field in \ref{fig:2}(b) at $t = 1.0$ ns is about $ 7.6 $ $\rm T \cdot mm$.

\subsection*{Hybrid Collisional-Collisionless Magnetic Reconnection: Experimental Results}

The proton radiography results of the magnetic field topologies at various times in Figs. \ref{fig:3}(a)-\ref{fig:3}(d) provide, for the first time, direct picture of the whole process for the hybrid collisional-collisionless magnetic reconnection. At time $t=0.3$ ns [\ref{fig:3}(a)], we see that the toroidal Biermann magnetic field still has not sufficiently developed and is not prominent, where some small-scale jet-like proton accumulations exist possibly due to the filamentary magnetic field arising from escaping of hot electrons\cite{Graziani2017}. At time $1.0$ ns [\ref{fig:3}(b)], two low-dose rings (marked by the blue dashed lines) form at respectively collisionless CH and collisional Cu plasmas, which indicate the toroidal Biermann magnetic fields form, similar to those in Figs. \ref{fig:2}(a) and \ref{fig:2}(b). We also see that the two magnetic fields already start to touch each other, consistent with the above estimations of their expansion velocities. Further, we see that more protons accumulate inside the Cu plasma bubbles than that of CH, indicating a stronger self-generated magnetic field, which is attributed to a steeper temperature gradient $\nabla T_e$ of Cu plasmas, see Supplementary Fig. \textcolor{blue}{\bf S2}. Also, the sharper proton accumulation radiography in CH plasmas may benefit from the uniform drive laser focal spot in our experiment, rather than the Gaussian one in Cu plasmas. 

\begin{figure*}
    \includegraphics[width= 14 cm]{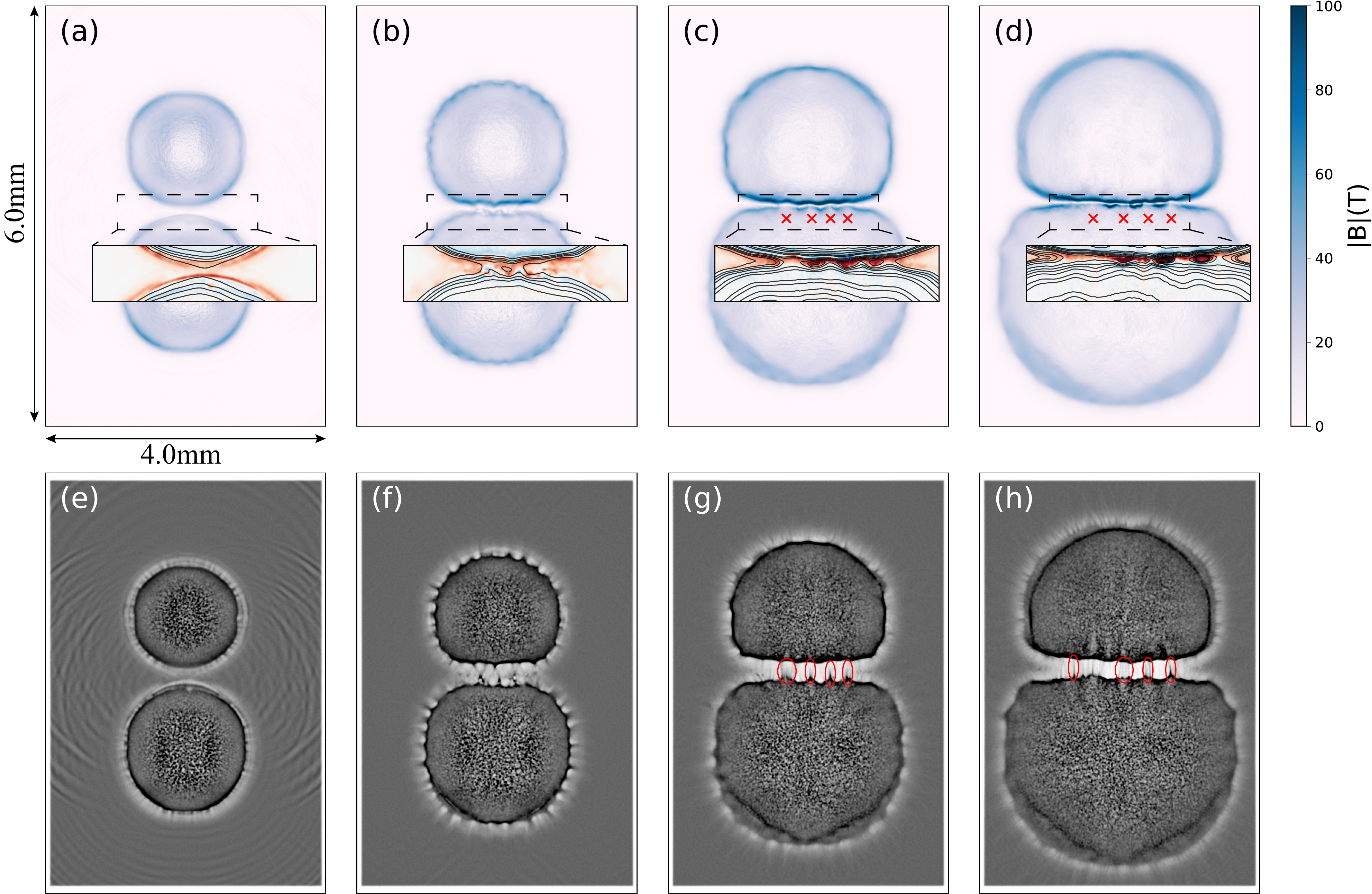}
    \caption{{\bf Kinetic 2D3V PIC simulation results of the hybrid magnetic reconnection.} Here the initial conditions are taken from the RMHD simulation at $z = 200 \ \mu$m, and at $t = 0.8$ ns [see Supplementary Fig. \textcolor{blue}{\bf S2}]. (a)-(d) are distributions of the magnetic field strengths $|B|$ (in blue color) at time $t = $ 1.0, 1.2, 1.5, 1.8 ns respectively during the magnetic reconnection. The zoomed images show the current densities $\textbf{J}$ (in blue-red color) and vector potential $A_z$ (contours) of the reconnection region, corresponding to the dashed box regions in (a)-(d) respectively. (e)-(h) show the corresponding synthetic proton radiography images based on the magnetic fields [(a)-(d)] obtained from PIC simulations (more details see Methods), where the higher the grayscale represents the higher the proton doses. The radiography geometric magnification $A$ and proton energy $E_p$ are consistent with the realistic experimental radiography setup. In order to match the extracted path-integrated magnetic fields $\psi$, the thickness of the fields is taken as $d \sim 100 \ \mu$m.}
    \label{fig:4}
\end{figure*}

At later time $t=1.5$ ns [\ref{fig:3}(c)], both Cu and CH plasmas continue to flow in, and as a result, the anti-parallel magnetic fields are strongly squeezed in their colliding region forming a reconnection current layer. Correspondingly, the proton radiography image [\ref{fig:3}(c)] shows that two low-dose rings merge into a long and narrow ribbon with length $\Delta x_{\rm exp} \sim 1.9$ mm and width $\Delta y_{\rm exp} \sim 0.3$ mm. Almost all protons are deflected out of this narrow ribbon region indicating that rather strong magnetic field exist locally, and the Y-shaped opening ends at both side of the ribbon region imply that the magnetic fields in Cu and CH plasmas begin to separate \cite{Tubman2021}. Finally and more importantly at time $t=1.8$ ns [\ref{fig:3}(d)], we see that the uniform low-dose ribbon is disturbed, and several high-dose fine filaments (marked by the red circles) appear, which means no proton deflection occurs there. In other words, it means that the magnetic fields are dissipated at these fine filamentary region, which is the key evidence for occurrence of the hybrid magnetic reconnection. 

To more accurately reflect the change of the magnetic field topology, the reconstructed path-integrated magnetic field $\psi$ and the corresponding vector potential $A_z$ (see field lines) are shown in Figs. \ref{fig:3}(f)-(h), see Methods. They clearly show that the two toroidal magnetic fields [see \ref{fig:3}(f)] are piled up together, forming a narrow reconnection layer dominated by only the $B_x$ component [see \ref{fig:3}(g)]. As shown in Fig. \ref{fig:3}(e) for the profile of $\psi$ along y-axis, we see that the asymmetry of the plasma properties results in the asymmetric magnetic flux (blue curve), which is stronger in Cu ($y > 0$) than CH ($y < 0$). And at later time, the current sheet on the side of the Cu plasma is obviously widened than that in CH plasmas (red curve), which may be the result of the magnetic diffusion caused by the strong collision effect. Using the Ampere's law $\textbf{J} = \nabla \times \textbf{B}$, it is estimated that the width of the current sheet (peak to peak) is about $2\delta_{\rm exp} \sim 150 \ \mu$m, which is much smaller than the length $2L \sim 1.9$ mm. For such a long and narrow current sheet, the tearing instability occurs and develops rapidly, resulting in formation of multiple plasmoids \cite{Uzdensky2016, Samtaney2009, Uzdensky2010}, which are shown clearly in \ref{fig:3}(h) by the red contours, also corresponding to the high-dose filaments in the radiography image \ref{fig:3}(d). Further, we see that there are still lots of un-reconnected magnetic fluxes on the outside of the fragmented current sheet, which may be due to the inefficient plasmoid ejection (because of high $\beta$) prevents further inflow of upstream magnetic fields \cite{Dong2012, Hare2017}. In view of the above radiography results, we conclude that, we have, for the first time, directly observed the growth of the tearing instability and formation of multiple plasmoids as well as the whole reconnection dynamics in the plasmoid-dominated magnetic reconnection.

\subsection*{Hybrid Collisional-Collisionless Magnetic Reconnection: Simulation Results}

\begin{table*}
\centering
    \caption{\label{tab:journals} {\bf Plasma parameters obtained from RMHD and following self-consistent PIC simulations}. (a) The expanding Cu and CH plasma parameters at $t=0.9$ ns obtained from 3D RMHD simulations, which are used as the initial condition for the following PIC simulations after self-similar scaling. (b) The plasma parameters inside the compressed reconnection layer (current sheet) of respectively Cu and CH sides at $t = 1.3$ ns obtained from PIC simulations, which are exactly the plasma states for magnetic reconnection. The listed plasma parameters include electron density $n_e$, electron temperature $T_e$, ion charge state $\bar{Z}$, magnetic field $B$, plasma flow velocity $v$ and Alfv\'en speed $v_A$, the ratio of thermal pressure to magnetic pressure $\beta$, the ratio of ram pressure to magnetic pressure $\beta_{\rm ram}$, Reynolds number $R_e$, magnetic Reynolds number $R_m$ and Peclet number $P_e$, electron mean free path $\lambda_{ei}$, Lundquist number $S$, current sheet width of the Sweet-Parker model $\delta_{\rm sp}$, ion skin depth $d_i$ and ion cyclotron radius $\rho_{ci}$.}
\begin{ruledtabular}
\begin{tabular}{lllllllllll}

\multicolumn{11}{c}{\bf (a) plasma parameters obtained from 3D RMHD simulations}\\
 \textbf{para}  & $ n_{\rm e} ({\rm 10^{18} cm^{-3}}) $ & $ T_{\rm e} ({\rm keV}) $ &  $\bar{Z}$   & $ B ({\rm T}) $  & $ v ({\rm km/s})$  & $ \beta $  & $ \beta_{\rm ram} $ &  $ R_{\rm e} $  & $  R_{\rm m} $  & $ P_{\rm e} $ \\
    \itshape {\rm Cu}\footnotemark[1]  & $2.5 \pm 1.0$  & $1.5 \pm 0.2$ & $20 \pm 2$  & $40 \pm 10$  &  $800 \pm 100 $  &  $1.0 \pm 0.2$ &  $6.7 \pm 1.0$  & $8000 \pm 2000$  & $3000 \pm 500 $  & $0.03 \pm 0.02$  \\
    \itshape {\rm CH}\footnotemark[1]  & $8.0 \pm 3.0$  & $1.1 \pm 0.1$ & $  3.5   $  & $25 \pm 10$  &  $1000 \pm 100$  &  $5.7 \pm 1.0$ &  $50  \pm 10$   & $120 \pm 30$ & $12000 \pm 3000$  & $0.05 \pm 0.03$  \\
\\    
 \multicolumn{11}{c}{\bf (b) plasma parameters in the reconnection layer obtained from PIC simulations}\\   
    \textbf{para}  & $ n_{\rm e} ({\rm 10^{18} cm^{-3}}) $ & $ T_{\rm e} ({\rm keV}) $ & $ B ({\rm T})$ & $ v_{\rm A} ({\rm km/s}) $  &  $\lambda_{\rm ei} ({\rm \mu m}) $ & $ \beta $ & $ S $ & $\delta_{\rm sp} ({\rm \mu m}) $ &  $ d_{\rm i} ({\rm \mu m}) $ & $\rho_{\rm ci} ({\rm \mu m})$  \\
    \itshape {\rm Cu}\footnotemark[2]  & $30 \pm 15$      & $2.5 \pm 0.5$   & $100 \pm 20$   & $200 \pm 50$ & $100 \pm 40 $  &  $4.0 \pm 2.0$ &  $1500 \pm 500$  & $25 \pm 10$ & $60 \pm 10$  & $20 \pm 5$ \\
    \itshape {\rm CH}\footnotemark[2]  & $40 \pm 20$      & $2.5 \pm 0.5$   & $50  \pm 10$   & $130 \pm 30$ & $500 \pm 200$  &  $16  \pm 5.0$ &  $5000 \pm 1500$ & $15 \pm 5$  & $50 \pm 10$  & $70 \pm 20$ \\
\end{tabular}
\end{ruledtabular} 
\footnotetext[1]{Plasma states as initial condition for PIC simulations. The plasma scale is about 1.0 mm and the Coulomb logarithm $\rm ln \Lambda$ is about 10.}
\footnotetext[2]{Plasma states for hybrid magnetic reconnection. The half length of the current sheet is estimated to be $L \sim 1.0$ mm.}
\label{tab:1}
\end{table*}

The above experimental results are reproduced and explained by a self-consistent combination of RMHD and PIC simulations as well as the proton radiography iterative inversion algorithm. As mentioned, magnetic reconnection is a rapid process of global magnetic topological self-organization triggered by local reconnection points, which depends critically on the plasma states. Before two plasma interactions, the laser-ablated plasma expansion dynamics is majorly governed by RMHD. Therefore, we firstly run the RMHD simulations with the ``FLASH'' code \cite{Fryxell2000} to obtain the basic plasma parameters as the initial condition for the following PIC simulations that can describe the interaction and reconnection dynamics of two plasmas. The expanding Cu and CH plasma parameters obtained from RMHD simulations are summarized in Table \ref{tab:1}(a), where the electron density $n_e$, temperature $T_e$, flow velocity $v$ and magnetic field strength $B$ are all taken for those at the periphery of each plasma bubble. All of them are comparable to the experimental results shown in the above Fig. \ref{fig:2}. More details can also be seen from the Methods and Supplementary Fig. \textcolor{blue}{\bf S2}. The differences in the equation of state (EOS) \cite{Heltemes2012} and opacity\cite{Macfarlane1989} of the Cu and CH lead to the asymmetric plasma states on both sides. For both plasmas, the thermal pressure to magnetic pressure ratio $\beta > 1$ and the ram pressure to magnetic pressure ratio $\beta_{\rm ram} \gg 1$ indicate that reconnection is strongly driven\cite{Rosenberg2015a, Rosenberg2015b}.

For the following plasma colliding and reconnection process, due to low densities and high temperatures of plasmas at their interaction region, the electron mean free path $\lambda_{\rm ei}$ comparable to the plasma scale $L$, therefore, the kinetic effects start to play key roles and kinetic PIC simulations are required \cite{Fox2011, Totorica2016, Xu2016}. In order to connect the PIC simulation with the RMHD simulation in a self-consistent manner, a self-similar transformation is proposed and applied here. Just like the self-similarity principle of the magnetohydrodynamic equations \cite{Remington1999, Ryutov2000, Ryutov2001}, we choose a set of free parameters to spatially scale down the plasma parameters obtained from the above RMHD simulation to the relative small scale that is practical for kinetic PIC simulation, where the plasma properties including $\beta$, $\beta_{\rm ram}$, the Mach number and the Alfv\'enic Mach number, etc. are all kept conserved between RMHD and PIC. The Coulomb collisions\cite{Nanbu1998} are also well included by making $L/\lambda_{ei}$ conserved. Through this novel methodology, see more details in Methods, we carry out the whole simulation in a self-consistent manner. 

\begin{figure*}
    \includegraphics[width= 15cm]{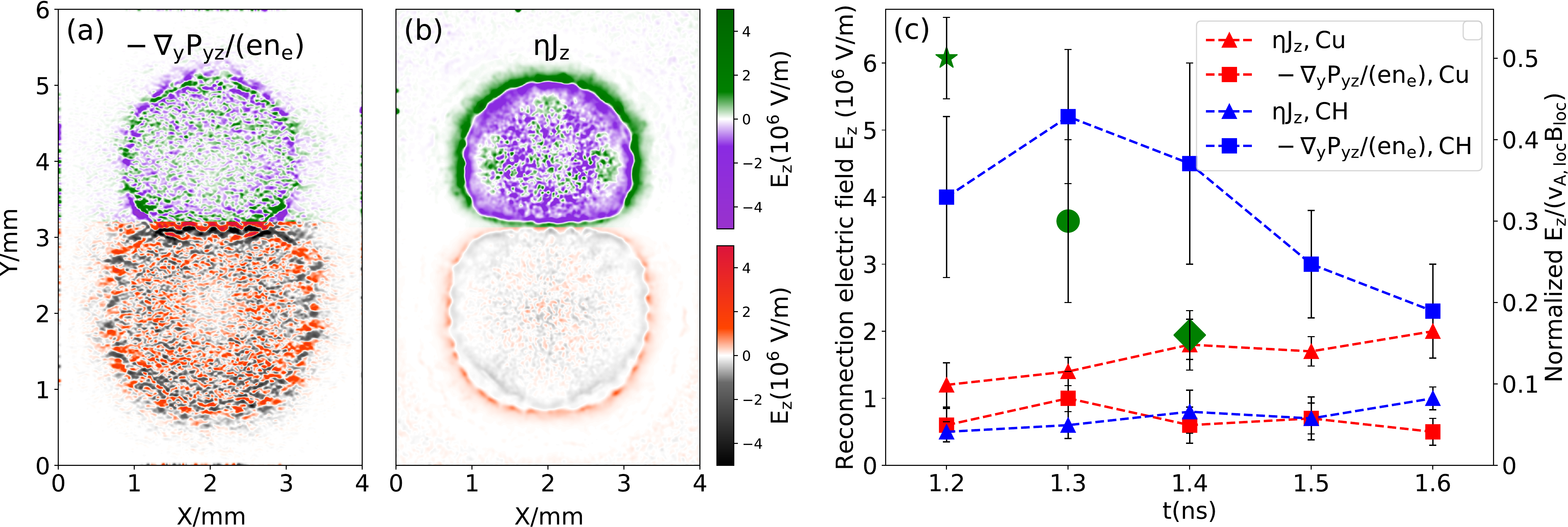}
    \caption{{\bf Contribution terms of the reconnection electric field $\mathbf E_{\bf z}$ in the hybrid magnetic reconnection.} (a) and (b) are respectively the contributions from the off-diagonal (non-gyrotropic) component of the electron pressure tensor $E_{z,p_e} \sim - \nabla_y P_{yz} / (e n_e)$ (using the colormap scale from black to red colors) and the resistivity $E_{z,\eta} = \eta J_z$ at $ t = 1.4 $ ns (using the colormap scale from purple to green colors), where, the same as the setup in experiment, the upper half part is collisional Cu plasma and the lower half part is collisionless CH plasma. (c) shows their contributions varying with time from $t=1.2$ to $1.6$ ns for respectively Cu and CH plasmas during reconnection. The normalized reconnection electric field $E_z$ (normalized to local Alfv\'en speed $v_{A,loc}$ and magnetic field $B_{loc}$) for respectively the cases of purely collisionless CH-CH (green star), hybrid Cu-CH (green circle) and purely collisional Cu-Cu (green diamond) reconnections are also marked in (c), where all the values are chosen at the time when the changes of their magnetic field topologies are most significant.}
    \label{fig:5}
\end{figure*}

The fully-kinetic PIC simulations are performed in two-dimensional (2D) xy plane with the code ``EPOCH'' \cite{Arber2015}. It has been identified that the 3D geometric effect has little effect on the reconnection rate in such a strongly-driven regime \cite{Rosenberg2015b}. Topologies of magnetic field evolutions at various times in the simulations are shown in Figs. \ref{fig:4}(a) to \ref{fig:4}(d) from $t=1.0$ to $1.8$ ns. Based on these field topologies, we also carry out the simulation for the synthetic proton radiography process, see Methods. The synthetic proton radiography images corresponding to \ref{fig:4}(a) to \ref{fig:4}(d) are shown respectively in Figs. \ref{fig:4}(e) to \ref{fig:4}(h). We see that the Cu and CH plasma bubbles start to contact with each other at $t=1.0$ ns [\ref{fig:4}(a)], in consistence with the experimental result [see \ref{fig:3}(b)].  The corresponding synthetic radiography image in \ref{fig:4}(e) also show similar feature as that in the above \ref{fig:3}(b), where the slight difference may be due to lack of a driving source. At $t=1.2$ ns [\ref{fig:4}(b)], similar as the experimental results, we see that the magnetic fields are compressed and amplified in the colliding region, forming a long and narrow current sheet. Due to the plasma pressure asymmetry, we also see the current sheet drifts slowly towards the Cu plasma side. 

Subsequently at time $t=1.5$ ns, we see clearly from \ref{fig:4}(c) that the current sheet breaks and multiple plasmoids form due to the tearing instability \cite{Uzdensky2016, Samtaney2009, Uzdensky2010}, and magnetic field energy dissipates at several local points (X-points), marked by the red crosses. Furthermore, we see all the plasmoids protrude into the CH plasma side and are more pronounced in the CH plasma side, which further verifying that the asymmetry of the tearing instability and magnetic reconnection. This feature is quite similar to the astronomical observations of the reconnection between solar filaments and coronal loops \cite{Li2016}. Again, both the X-points and plasmoids can be accurately reflected by the synthetic proton radiography, as shown in \ref{fig:4}(g), where several high-dose fine filaments appear (marked by the red circles), agreeing well with the experimental results in Fig. \ref{fig:3}(c). Note that the length and width of the low-dose ribbon in \ref{fig:4}(g) is about $\Delta x_{\rm sim} \sim 1.7$ mm and $\Delta y_{\rm sim} \sim 0.25$ mm, which are also both consistent with the experimental results. At much later time $t=1.8$ ns [\ref{fig:4}(d)], we see that the magnetic fields, current sheet and plasmoids are further continuously compressed, and the distance between these plasmoids gradually increases due to stretch and ejection \cite{Dong2012, Hare2017} ($v \sim 10^5 \ {\rm m/s}$). These plasmoids and X-points become more prominent in radiographic images, see \ref{fig:4}(h) and also \ref{fig:3}(d). The number of plasmoids and X-points does not change with time, implying that the merging of magnetic islands and secondary tearing instability \cite{Shibata2001, Shibata2016, Uzdensky2016, Samtaney2009, Uzdensky2010} both do not occur here. The phenomenological consistency between the simulation and the experimental results verifies that the tearing instability and plasmoid-dominated magnetic reconnection indeed take place in our experiment.  
  
To reveal the inherent mechanism of the hybrid collisional-collisionless magnetic reconnection, the main parameters of both Cu and CH plasmas inside the reconnection layer obtained from PIC simulations are summarized in Table \ref{tab:1}(b), which are exactly the plasma states for magnetic reconnection. We see that since both Cu and CH plasmas are highly compressed, their densities and temperatures significantly increase, and the magnetic field strengths are amplified to more than 2 times than their initial values, see also Supplementary Fig. \textcolor{blue}{\bf S3}. The average width of the reconnection layer (current sheet) is about $2\delta_{\rm sim} \sim 100 \ \mu$m, which is similar with the reconstructed experimental result $2\delta_{\rm exp} \sim 150 \ \mu$m. The Lundquist number $S \equiv L v_A / \eta \gg 1$ in both plasmas, especially in the CH plasma where $S \sim 0.5 \times 10^4$, verifying that the tearing instability can easily occur and develop. The half width of the Sweet-Parker current sheet $\delta_{\rm sp} \equiv L / \sqrt{S}$ of Cu (about $0.4d_i$) and CH (about $0.3d_i$) plasmas are both smaller than the ion skin depth $d_i$ or ion cyclotron radius $\rho_{ci}$, which means the two-fluid effect \cite{Tubman2021, Rosenberg2015b} needs to be considered. More importantly, we see that, due to the very different ion charge states $\bar{Z}$ of Cu ($\bar{Z}\approx20$) and CH ($\bar{Z}\approx3.5$), the plasma on the Cu side is in a strongly-collisional state ($\lambda_{\rm ei} \sim 100 \ \mu$m), while the plasma on the CH side is collisionless ($\lambda_{\rm ei} \sim 500 \ \mu$m). That is, an asymmetric magnetic reconnection in the hybrid collisional-collisionless regime is formed.

From both the above experimental and simulation results [comparing Figs. \ref{fig:3}(c) and \ref{fig:3}(d), or comparing \ref{fig:4}(c) and \ref{fig:4}(d)], we estimate that the characteristic growth time of the tearing instability for the hybrid reconnection is rather fast as about $\tau_{\rm TI}=0.5$ ns, at the same order of the Alfv\'en transit time across the current sheet $\tau_{A} \equiv \delta / v_A$. Such fast tearing instability and reconnection dynamics cannot be explained by only the two-fluid effect. This can be proved from two aspects. One the one hand, as known, the maximum growth rate of the spontaneous tearing instability from the hydrodynamic perspective is about $\gamma_{\rm max} \sim (v_{A}/\delta) (d_e/d_i)^{0.7}(d_i/\delta)^{1.5}$ \cite{Fitzpatrick2004, Loureiro2017}. Substituting the reconnection plasma parameter in Table \ref{tab:1}(b) into it and assuming $\delta \sim 50 \ \mu$m, we estimate that $\gamma_{\rm Cu} \sim (2.5 \pm 1.0) \times 10^{8} \ {\rm s}^{-1}$ and $\gamma_{\rm CH} \sim (1.5 \pm 0.5) \times 10^{8} \ {\rm s}^{-1}$, which corresponds to the characteristic growth time of the tearing instability as both larger than $10$ ns, too slow to explain the experimental and simulation results. On the other hand, considering only the two-fluid effect, the predicted most unstable mode is $k_{\rm max} \sim (1/\delta)(d_e/d_i)^{0.3}(d_i/\delta)^{0.5}$\cite{Fitzpatrick2004, Loureiro2017}, which means that the most confident number of X-points in reconnection is $M = L k_{\rm max}/\pi \sim 2$, also smaller than 4 observed in our experiment [see \ref{fig:3}(c)].

Therefore, to explain the fast and effective tearing instability observed in our experiment, we analyze the various contributions of the dissipative reconnection electric field $E_z$, which after normalization is defined as the reconnection rate. The two main contributions, the non-gyrotropic component of the electron pressure tensor $E_{z,p_e} \sim - \nabla_y P_{yz} / (e n_e)$ and the resistive component $E_{z,\eta} = \eta J_z$, are shown in Fig. \ref{fig:5}(a) and \ref{fig:5}(b) respectively at time $t=1.4$ ns. Their contributions varying with time from $t=1.2$ to $1.6$ ns for respectively Cu and CH plasmas during reconnection are also shown in Fig. \ref{fig:5}(c). We see clearly that the non-gyrotropic electron pressure tensor always dominates [the lower half part in \ref{fig:5}(a) and the blue square symbol in \ref{fig:5}(c)] in the collisionless CH plasma side, while the resistive electric field dominates [the upper half part in \ref{fig:5}(b) and the red triagnle symbol in \ref{fig:5}(c)] in the collisional Cu plasma side. And as a whole, for the hybrid reconnection, the collisionless non-gyrotropic electron pressure tensor contributes (about $75\%$) most of the reconnection electric field, while the resistivity only contributes $\sim 25\%$. In other words, it is just the collisionless non-gyrotropic electron pressure that results in the fast tearing instability and reconnection in our experiment \cite{Hosseinpour2014}. The total reconnection electric field is about $E_z \sim 6 \times 10^6 \ \rm{V/m}$, on the order of $ 0.3 \pm 0.1 \  v_{A,loc} B_{loc} $ after normalization [marked by the green circle in \ref{fig:5}(c)], i.e., a fast reconnection rate \cite{Fox2011, Xu2016}.
   
To further demonstrate the dynamics of our hybrid reconnection, we also carry out PIC simulations for the cases of the purely collisionless CH-CH and purely collisional Cu-Cu magnetic reconnection, where the drive laser and other parameters are also the same. The results are shown in Supplementary Fig. \textcolor{blue}{\bf S4}. We see that in the purely collisionless reconnection case, the tearing instability grows even more quickly, resulting in formation of more number of multiple X-points, where the normalized reconnection electric field is $E_z \sim 0.5 \pm 0.05 \  v_{A,loc} B_{loc}$ [marked by the green star in \ref{fig:5}(c)], a little higher than the hybrid case. By contrast, in the purely collisional Cu-Cu reconnection case, the resistive electric field dominates, a Sweet-Parker-like current sheet with single X-point is formed, where $E_z \sim 0.16 \pm 0.03 \  v_{A,loc} B_{loc}$ [marked by the green diamond in \ref{fig:5}(c)] is much lower. These further confirms the occurrence of a hybrid collisional-collisionless magnetic reconnection in our experiment. 

\section*{Discussion}

In summary, we have carried out the first experiment on an unexplored new regime of magnetic reconnection, hybrid collisional-collisionless regime, by colliding of laser-ablated high-Z Cu and low-Z CH plasmas. Using highly-resolved proton radiography, we have, for the first time, directly observed the topology changes for the whole process of the plasmoid-dominated magnetic reconnection. We find that the growth rate of the tearing instability in such a hybrid regime is still extremely large, resulting in rapid formation of multiple plasmoids, lower than the purely collisionless case but much higher than the collisional case. With the self-consistent simulations, we show that, in this hybrid magnetic reconnection, the reconnection electric field show much distinct feature, which is large and grows fast at the collisionless plasma side induced by the off-diagonal (non-gyrotropic) component of the electron pressure tensor, whereas smaller and more slowly at the collisional side induced by only the resistivity. 

The hybrid collisional-collisionless magnetic reconnection discussed here occurs widely in astrophysics. For example, for the reconnection between the erupting collisional filaments and the collisionless coronal loops in solar and stellar atmosphere, the astronomical observations reveal that most of the plasmoid formation and plasma heating occurred on the side of the collisionless coronal loops \cite{Li2016}, in consistence with our results. Besides, the possible reconnections occurring between the dense accretion disk and the tenuous interstellar medium may also be in such hybrid regime. Our experimental results show that the plasmoid formation is a universal feature of magnetic reconnection, even in this hybrid regime. We also show that the specific magnetic reconnection mechanism depends critically on the plasma collision rates, which significantly affect the growth of the tearing instability and further the global reconnection rate, eventually determining the energy conversion efficiency from fields to plasmas, such as heating the solar corona \cite{Schrijver1998}. Moreover, the lower plasma $\beta$, higher Lundquist number $S$ and ubiquitous guiding fields in these astrophysical environments may promote the plasmoid ejection, thereby help to establish the positive feedback for global fast reconnection \cite{Shibata2001, Shibata2016}.

\section*{Methods}
\subsection*{Experimental setup}

The hybrid collisional-collision magnetic reconnection experiments presented in this work are carried out on ShenGuang II upgrade (SG-II-U) laser facility at the Joint Laboratory on High Power Laser and Physics, Shanghai Institute of Optics and Fine Mechanics, Chinese Academy of Sciences. The laser facility have 8 nanosecond (ns) pulses and 1 picosecond (ps) petawatt pulse. The main target consists of two foils with the upper half part as Cu foil of density $\rho = 8.92$ g/cm$^3$ and the lower half part as plastic CH foil of density $\rho = 1.02$ g/cm$^3$. Both foil targets have thickness of 30 $\mu$m and width of 1.0 mm, and parallel to each other, and there is a 0.5 mm gap between the Cu and CH targets, in order to eliminate the scattering effect of protons by the solid targets. 

The two foil targets are ablated with two $3 \omega$ long laser pulses (green) that have energy about $800 \pm 100 $J in a $ 1 \pm 0.1 $ ns square temporal profile, with either a 200 $\mu$m FWHM Gaussian or a 450 $\mu$m diameter uniform focal spot. The corresponding average laser intensities are about $2.5 \times 10^{15}$ W/cm$^2$ and $4.0 \times 10^{14}$ W/cm$^2$ respectively. The incident angle between the laser beam and the target normal is about 45$^\circ$, and the distance between the two focal spots is about 1.8 mm. The proton radiography is set up along the face-on ($-z$-axis) direction, by irradiating the backlighter 20-$\mu$m-thick Ta foil (red) target with a relativistic ps pulse. The ps pulse has energy of $ 350 \pm 50 $ J, wavelength of 1.053 $\mu$m, duration of $1.0 \pm 0.1$ ps and is focused by a 800 mm focal-length, $f /2.5$ off-axis parabolic mirror to an intensity about $3.0 \times 10^{19}$ W/cm$^2$, with focal spot diameter $\sim 40 \mu$m. The cut-off energy of the high-quality laminar proton beam exceeds 30 MeV, and the deflected protons are finally deposited on the HD-V2 radiochromic film (RCF) stack, whose response to proton dose has been well calibrated with a good linear relationship between proton dose and optical density (OD). The distance between the backlighter foil (Ta) and the main foils (Cu and CH) is $h \sim 10$ mm, and that between the main foils and the RCF stack is $ L \sim 50 $ mm, which result in the geometric magnification factor of the proton radiography system as about $ M \equiv (h+L)/h \sim 6.0 $.

\subsection*{Path-integrated field reconstruction}
The extraction of the path-integrated magnetic field $\psi$ from proton radiographs in this work is done by using the algorithm and program ``PROBLEM'' \cite{Bott2017}, where a parabolic Monge-Amp\`ere equation is solved. Note that the formation of the proton radiographys can be understanding as following.

First, the propagation of laser-produced proton beam initially can be regarded as a uniform expansion from a point source (laser-driven target normal sheath acceleration region) into the probing HED plasma region. Under the paraxial approximation, $ l_p / r_i \ll 1 $, when the proton beam arrives at the probing region, it can be regarded as a two-dimensional near-planar sheet, where $ l_p $ and $ r_i $ are respectively the plasma lateral length and the distance from the proton source to the plasma, in our experiments, $l_p \approx 1$ mm and $ r_i  = h \approx 10 $ mm.

Afterwards, the proton beam is deflected by the Lorentz force arising from self-generated magnetic fields in HED plasmas, and the lateral deflection velocity can be expressed as
\begin{equation}
  \textbf{v}(\textbf{x}_{\perp 0}) \approx \frac{e}{m_p c} \hat{\textbf{z}} \times \int_{0}^{l_z} \mathrm{d}s \textbf{B}(\textbf{x}(s)),
\end{equation}
where $e$ is the proton charge, $m_p$ is the proton mass, $\textbf{x}_{\perp 0}$ and $\textbf{x}(s)$ are respectively the initial perpendicular position and the proton trajectory. The outgoing proton beam moves approximately in a straight line at a constant velocity, and the lateral deflection velocity $ \textbf{v}(\textbf{x}_{\perp 0}) $ is amplified as
\begin{equation}
   \textbf{x}_{\perp}^{(s)}(\textbf{x}_{\perp 0}) \approx \frac{r_i + r_s}{r_i} \textbf{x}_{\perp 0} + \frac{\textbf{v}(\textbf{x}_{\perp 0})}{v_p} r_s,\end{equation}
where $ v_p $ is the proton velocity, $ r_s $ is the distance from the plasma to the RCF screen, $r_s = L \approx 50$ mm in experiments, and $\textbf{x}_{\perp}^{(s)}$ is the proton position on the screen.

Eventually, the deflection effect by the magnetic field makes the initially uniform proton beam redistribute on the RCF screen as
 \begin{equation}
        \Psi(\textbf{x}_{\perp}^{(s)}(\textbf{x}_{\perp 0})) = \frac{1}{|\mathrm{Det}( \nabla_{\perp 0} [\textbf{x}_{\perp}^{(s)}(\textbf{x}_{\perp 0})] )|} \Psi_0,
\end{equation}
where $\Psi$ and $ \Psi_0 $ are respectively the redistributed proton flux on the RCF screen and the initial proton flux (generally uniform), and $ \nabla_{\perp 0} $ is the gradient operator of the initial plasma coordinates.

Therefore, the path-integrated magnetic field can be obtained by solving the Monge–Ampère equation (4), and
\begin{equation}
        \psi \equiv \int_{0}^{l_z} \mathrm{d}s \textbf{B}(\textbf{x}(s)) = - \frac{m_p c}{e} \hat{\textbf{z}} \times \textbf{v}(\textbf{x}_{\perp 0}).
\end{equation}

Due to the divergence-free of magnetic field $\nabla \cdot \textbf{B} = 0$, then the deflection velocity is curl-free $\nabla \times \textbf{v} = 0$. The deflection velocity can be viewed as the gradient of a potential function $\textbf{v} \equiv \nabla \phi$, and the corresponding magnetic field potential function is expressed as
\begin{equation}
    \mathcal{A} = - \frac{m_p c}{e} \phi,
\end{equation}
under the quasi-two-dimensional approximation, this is equivalent to the magnetic vector potential $A_z$, and $A_z \sim \mathcal{A} / l_z$.

This inversion problem is well-defined, the geometric distance and proton energy are determined in the experiments, the redistributed proton flux $\Psi$ can be obtained from the response function of RCF optical density (OD) to proton dose, and which has an almost linear relationship in the low-dose region. The inhomogeneity of the initial proton flux $ \Psi_0 $ is ignored, see also Fig. \textcolor{blue}{\bf S1} in Supplementary.

\subsection*{3D radiation-magnetohydrodynamic (RMHD) simulation}

The three-dimensional (3D) RMHD simulations in this work are carried out by using the code ``FLASH'' \cite{Fryxell2000}, which has been developed to include many high energy density physics modeling capabilities including laser energy deposition, multi-temperature ($ T_e \neq T_i \neq T_{rad} $),  anisotropic electron thermal conductivity, and multi-group radiation transport etc. We have also further extended the code with the self-consistent magnetic field modeling capabilities including the Biermann battery, Nernst advection and so on, which is the reason why we called it as "RMHD". 

The initial conditions of 3D RMHD simulations are taken following those in the experiments. The size of the simulation box is $4\times6\times1.5$ $\rm mm^3$ in all $(x, y, z)$ directions and the solid CH/Cu foil targets with thicknesses $\rm 30\ \mu m$ in z direction and areas of $1.0\times2.0$ $\rm mm^2$ in respectively x and y directions are placed at the position $z=0.2\ \rm{mm}$. The ns laser pulses with energy of 800J (a multiplier is used to account for the scattering caused by parameter instabilities and match the plasma morphology in the experiments), flat-top temporal profile of duration 0.9 ns plus 0.1ns rising and falling times, and focal spots consistent with experiments are incident on the foil targets at an oblique angle of about 45$^{\circ}$. The equation of state (EOS) and opacity of solid foil materials (CH and Cu) come from the code BADGER \cite{Heltemes2012} and IONMIX \cite{Macfarlane1989}, respectively. The initial densities of the solid foil targets are set to $\rho_{\rm CH} = 1.02$ g/cm$^3$ and $\rho_{\rm Cu} = 8.92$ g/cm$^3$, and the rest is filled with background He gas with a density of $2 \times 10^{-6}$ g/cm$^3$. All the initial temperatures are set to be uniform as room temperatures $T_0 \sim 290$ K.

In the simulations, an adaptive grid is used. And the geometric-optical approximation of inverse bremsstrahlung absorption is used to represent laser heating. The Riemann solver of the MHD equations uses HLLC, and the slope limiter uses minmod. The Courant (CFL) number is self-adaptive, the typical value is 0.4. Fully implicit electron thermal conductivity is also adopted, and the flux limiter is selected as a typical 0.06. The boundary conditions for fluid, electron thermal conductivity and radiation transport are all set to open. The simulation results of plasma properties are shown in Supplementary Fig. \textcolor{blue}{\bf S2}.

\subsection*{Self-consistent RMHD-PIC simulation}
The fully-kinetic PIC simulations in this work are performed in two-dimensional (2D) xy plane with the code ``EPOCH'' \cite{Arber2015}. It has been identified that the 3D geometric effect has little effect on the reconnection rate in such a strongly-driven regime \cite{Rosenberg2015b}. In order to connect the PIC simulation with the RMHD simulation in a self-consistent manner, a self-similar transformation is proposed and applied. That is, before the kinetic effects become important, RMHD is used to simulate the macroscopic states of the plasmas, and then PIC is used to simulate the subsequent kinetic processes, while ensuring a self-consistent connection between them. Just like the self-similarity principle of the magnetohydrodynamic equations \cite{Remington1999, Ryutov2000, Ryutov2001}, we choose a set of free parameters to spatially scale down the plasma parameters obtained from the above RMHD simulation to the relative small scale that is practical for kinetic PIC simulation, where the plasma properties including $\beta$, $\beta_{\rm ram}$, the Mach number and the Alfv\'enic Mach number, etc. are all kept conserved between RMHD and PIC. The Coulomb collisions are also well included by making $L/\lambda_{ei}$ conserved.

The principle of this method comes from the self-similarity of the ideal MHD equations \cite{Remington1999, Ryutov2000, Ryutov2001}, that is, under a set of specific parameter transformation, 
\begin{equation}
\begin{aligned}
    \textbf{r}_1 &= a \ \textbf{r}_0, \quad \rho_1 = b \ \rho_0, \quad p_1 = c \ p_0, \\ 
    \textbf{v}_1 = &\sqrt{c/b} \  \textbf{v}_0, \quad \textbf{B}_1 = \sqrt{c} \ \textbf{B}_0, \quad t_1 = a \sqrt{b/c} \ t_0,
\end{aligned}
\end{equation}
the form of the ideal MHD equations remains unchanged. $a$, $b$ and $c$ are the free transformation parameters. $\textbf{r}$, $\rho$, $p$, $\textbf{v}$, $\textbf{B}$ and $t$ respectively represent length, density, pressure, velocity, magnetic field and time. The subscripts 0 and 1 respectively represent two different systems connected by this parameter transformation. This means that under similar initial and boundary conditions, the two systems have the same evolution characteristics. This scale transformation is often used in laboratory astrophysics to compare experiment results with astrophysical events. 

The condition for the establishment of this self-similarity is that the plasmas is polytropic and ideal, which means that viscosity, electric resistivity and heat conductivity are not dominant, i.e., the Reynolds number $R_e \equiv L v / \nu $, the magnetic Reynolds number $R_m \equiv L v / \eta $ and the Peclet number $P_e \equiv L v / \chi $ are far greater than 1.0. These dimensionless numbers, before the Cu and CH plasma bubbles touch each other, are listed in Table \ref{tab:1}(a). Both $R_e$ and $R_m$ are far greater than 1.0. Although $P_e$ is less than 1.0, considering that the plasma corona is approximately isothermal (see Supplementary Fig. \textcolor{blue}{\bf S2}) and the suppression of the electron heat flow by the magnetic fields, the coronal $P_e$ will be greater than 1.0. And the coronal plasma is approximately an ideal gas, so it is also polytropic. Therefore, under limited approximations, the laser-driven expanding plasmas can be regarded as an ideal polytropic gas, where the self-similar transformation can be applied.

In the context of PIC simulations, since the plasma density has the relation $\rho \sim m_i n_i$ and pressure $p \sim n T$, the transformation parameters should be chosen as $b = f_m f_n$ and $c = f_n f_T$, where $f_m$, $f_n$ and $f_T$ are the transformation parameters of ion mass, plasma density and plasma temperature respectively. As for the transformation parameter $a$ of length, in order to correctly account for the two-fluid effect, it should be ensured that $L/d_i$ (on the order of 100 in typical laser-ablated plasmas) remains unchanged under the similar transformations, and this constraint gives $a = \sqrt{f_m/f_n}$, which also maintains compatibility with the Ampere’s law $\textbf{J} = \nabla \times \textbf{B}$. In this way, the self-similar transformation of ideal RMHD equations is self-consistently converted into the language of PIC simulation, that is, the transformation parameters $a$, $b$ and $c$ of RMHD variables are converted into $f_m$, $f_n$ and $f_T$ of PIC particle variables,
\begin{equation}
    a = \sqrt{f_m/f_n}, \quad b = f_m f_n, \quad c = f_n f_T.
\end{equation}
Other variables are also transformed according to the corresponding parameters, the fluid velocity $\textbf{v}_1 / \textbf{v}_0 = \sqrt{f_T / f_m} = c_{s,1} / c_{s,0}$, i.e., the ratio of the ion sound velocity of the two systems. And the time $ t_1 / t_0 = f_m/\sqrt{f_n f_T} = f_m / \sqrt{c} = \omega_{ci,1}^{-1} / \omega_{ci,0}^{-1}$, the ratio of the ion cyclotron time of the two systems.

We apply this novel transformation methodology to the kinetic simulation of hybrid magnetic reconnection, and transform the results of the RMHD simulation at $t = 0.8$ ns to the PIC simulation, see Supplementary Fig. \textcolor{blue}{\bf S2} and Fig. \textcolor{blue}{\bf S3}. Balance the computing power and the universality, we choose the mass of Cu ions in the PIC simulation to be $m_{\rm Cu} = 1000 m_e$, and the mass of CH ions $m_{\rm CH}$ and He ions $m_{\rm He}$ are calculated proportionally, which gives $f_m \sim 0.0085$. The parameters of number density $f_n = 10$ and temperature $f_T = 5$. These parameters lead to $t_1 / t_0 \sim 0.0012$, i.e., 1.2 ps in PIC simulations represents 1.0 ns in RMHD simulations and experiments. In the PIC simulations, the ion charge state $\bar{Z}$ is always consistent with the RMHD simulations. The whole simulation box is divided into $1600 \times 2400$ grids, and 100 macro-particles per species are employed in each grid. Open boundary conditions are also applied in each direction. Other set of transformation parameters hardly affect the final results.

Although the introduction of collision effects will destroy the self-consistency of similar transformations (due to the bad-scaled transport coefficient $\nu$, $\eta$ and $\chi$), as long as the dimensionless parameters $R_e$, $R_m$ and $P_e$ are far greater than 1.0, then this deviation is negligible. In the hybrid collisional-collisionless magnetic reconnection, the collision effect of Cu plasma may be important near the current sheet, where $\lambda_{ei} \sim 2\delta$. The commonly used parameter to evaluate the strength of the collision effect is $L/\lambda_{ei}$. In the PIC simulation, we keep this parameter consistent with the RMHD simulation and experiments. Because of $\lambda_{ei} \propto T_e^2/n_e$, the transformation parameter of the collision frequency is $\nu_{ei,1} / \nu_{ei,0} = f_T^2 / \sqrt{f_m f_n}$. By this way, the binary Coulomb collision \cite{Nanbu1998} can be appropriately included.

\subsection*{Synthetic proton radiography simulation}

In order to directly compare with the experimental proton radiography images, we perform synthetic, numerical proton radiography simulation based on the magnetic fields obtained from the self-consistent combined MHD-PIC simulation. The main steps are as follows: protons with a point source are magnified by point projection, then enter the area filled with magnetic fields and deflect by the Lorentz force. After escaping these magnetic fields, they perform a uniform linear motion and are finally deposited in the simulated RCF stacks. These protons are regarded as PIC macro-particles, and their energy is consistent with the experiments. The geometry and magnification rate of the system are also consistent with the experiments. The deflection motion of these protons in the magnetic fields adopts the Boris algorithm, and the magnetic field fluxes are equivalent to the path-integrated magnetic field $\psi$ that extracted from the experimental images. These protons deposit energy in the simulated RCF stacks and gradually slow down until they stop. The stopping power of different materials come from the NIST Standard Reference Database \cite{NIST}, and the scattering effect is ignored. The accuracy of this code has been well benchmarked.

\section*{Data availability}

All data that support the findings of this study are available from the corresponding author upon reasonable request.

\section*{Code availability}

FLASH is  an open radiation MHD simulation code for plasma physics and astrophysics, developed by the DOE NNSA-ASC OASCR Flash Center at the University of Chicago. The code is available for download from \url{ http://flash.uchicago.edu/site/flashcode/}.

EPOCH is the Extendable PIC Open Collaboration project to develop a UK community advance relativistic EM PIC code. The open source code is available for download from \url{https://cfsa-pmw.warwick.ac.uk/EPOCH/epoch}.

PROBLEM solver is an open source Python implementation of the proton radiography reconstruction algorithm of Bott (2017), the source code is available for download from \url{ https://github.com/flash-center/PROBLEM}.

\section*{Acknowledgements}
  This work is supported by Science Challenge Project, No. TZ2018005; National Natural Science Foundation of China, Grant Nos. 12135001, 11825502 and 11921006; the Strategic Priority Research Program of Chinese Academy of Sciences, Grant No. XDA25050900; BQ acknowledges support from National Natural Science Funds for Distinguished Young Scholar, Grant No. 11825502. The simulations are carried out on the Tianhe-2 supercomputer at the National Supercomputer Center in Guangzhou.
  
\section*{Author contributions}
B.Q. and S.P.Z. proposed and were in charge of the research campaign as principle investigators. Z.H.Z., Y.X., Z.L., W.P.Y., W.Q.Y. and B.Q. carried the simulations, the data analysis. H.H.A., Z.H.Z., J.X., C.W., J.J.Y, Z.Y.X., Z.H.F., A.L.L., W.M.Z. and B.Q. carried out the experiments. B.Q., and Z.H.Z. wrote the paper. W.B.P. S.P.Z. and X.T.H. contribute to theoretical interpretations of the results.

\section*{Competing interests}
The authors declare no competing interests.

\section*{Additional information}
Correspondence and requests for materials should be addressed to B.Q.


\section*{References}
\begin{thebibliography}{99}
\scriptsize
    \bibitem{Zweibel2009}
        Zweibel, E. G. \& Yamada, M. Magnetic Reconnection in Astrophysical and Laboratory Plasmas. \emph{Annu. Rev. Astron. Astrophys.} {\bf 47}, 291–332 (2009).

    \bibitem{Yamada2010}
        Yamada, M., Kulsrud, R. \& Ji, H. Magnetic reconnection. \emph{Rev. Mod. Phys.} {\bf 82}, 603–664 (2010).

    \bibitem{Schrijver1998}
        Schrijver, C. J. \emph{et al.} Large-scale coronal heating by the small-scale magnetic field of the Sun. \emph{Nature} {\bf 394}, 152–154 (1998).

    \bibitem{Piran2004}
        Piran, T. The physics of gamma-ray bursts. \emph{Rev. Mod. Phys.} {\bf 76}, 1143–1210 (2004).

    \bibitem{Biskap1997}
        Biskamp. D. Collisional and collisionless magnetic reconnection. \emph{Phys. Plasmas} {\bf 4}, 1964–1968 (1997).

    \bibitem{Shibata2007}
        Shibata, K. Chromospheric Anemone Jets as Evidence of Ubiquitous Reconnection. \emph{Science}. {\bf 318}, 1591–1595 (2007).

    \bibitem{Sweet1958}
        Sweet, P. A. The Production of High Energy Particles in Solar Flares. \emph{Nuovo Cim} {\bf 8}, 188–196 (1958).

    \bibitem{Parker1963}
        Parker, E. N. The solar flare phenomenon and the theory of reconnection and annihilation of magnetic fields. \emph{Astrophys. J. Suppl. Ser.} {\bf 8} 177–211 (1963).

    \bibitem{Cassak2016}
        Cassak, P. A. \& Fuselier, S. A. Reconnection at Earth’s Dayside Magnetopause. in \emph{Magnetic Reconnection: Concepts and Applications} (eds. Gonzalez, W. \& Parker, E.) 213–276 (Springer International Publishing, 2016).

    \bibitem{Petrukovich2016}
        Petrukovich, A., Artemyev, A. \& Nakamura, R. Magnetotail Reconnection. in \emph{Magnetic Reconnection: Concepts and Applications} (eds. Gonzalez, W. \& Parker, E.) 277–313 (Springer International Publishing, 2016).

    \bibitem{Hesse2001}
        Hesse, M., Birn, J. \& Kuznetsova, M. Collisionless magnetic reconnection: Electron processes and transport modeling. \emph{J. Geophys. Res. Sp. Phys.} {\bf 106}, 3721–3735 (2001).

    \bibitem{Hesse2011}
        Hesse, M., Neukirch, T., Schindler, K., Kuznetsova, M. \& Zenitani, S. The Diffusion Region in Collisionless Magnetic Reconnection. \emph{Space Sci. Rev.} {\bf 160}, 3–23 (2011).

    \bibitem{Li2016}
        Li, L. \emph{et al.} Magnetic reconnection between a solar filament and nearby coronal loops.\emph{ Nat. Phys.} {\bf 12}, 847–851 (2016).

    \bibitem{Yang2019}
        Yang, B. \& Chen, H. Filament Eruption and Its Reformation Caused by Emerging Magnetic Flux. \emph{Astrophys. J.} {\bf 874}, 96 (2019).

    \bibitem{Nilson2006}
        Nilson, P. M. \emph{et al.} Magnetic reconnection and plasma dynamics in two-beam laser-solid interactions. \emph{Phys. Rev. Lett.} {\bf 97}, 255001 (2006).

    \bibitem{Li2007}
        Li, C. K. \emph{et al.} Observation of megagauss-field topology changes due to magnetic reconnection in laser-produced plasmas. \emph{Phys. Rev. Lett.} {\bf 99}, 055001 (2    007).

    \bibitem{Zhong2010}
        Zhong, J. \emph{et al.} Modelling loop-top X-ray source and reconnection outflows in solar flares with intense lasers. \emph{Nat. Phys.} {\bf 6}, 984–987 (2010).

    \bibitem{Fiksel2014}
        Fiksel, G. \emph{et al.} Magnetic reconnection between colliding magnetized laser-produced plasma plumes. \emph{Phys. Rev. Lett.} {\bf 113}, 105003 (2014).

    \bibitem{Rosenberg2015a}
        Rosenberg, M. J. \emph{et al.} Slowing of Magnetic Reconnection Concurrent with Weakening Plasma Inflows and Increasing Collisionality in Strongly Driven Laser-Plasma Experiments. \emph{Phys. Rev. Lett.} {\bf 114}, 205004 (2015).

    \bibitem{Rosenberg2015b}
        Rosenberg, M. J. \emph{et al.} A laboratory study of asymmetric magnetic reconnection in strongly driven plasmas. \emph{Nat. Commun.} {\bf 6}, 6190 (2015).

    \bibitem{Shibata2001}
        Shibata, K. \& Tanuma, S. Plasmoid-induced-reconnection and fractal reconnection. \emph{Earth Sp. Sci.} {\bf 53}, 473–482 (2001).

    \bibitem{Shibata2016}
        Shibata, K. \& Takasao, S. Fractal Reconnection in Solar and Stellar Environments. in \emph{Magnetic Reconnection: Concepts and Applications} (eds. Gonzalez, W. \& Parker, E.) 373–407 (Springer International Publishing, 2016).

    \bibitem{Uzdensky2016}
        Uzdensky, D. A. \& Loureiro, N. F. Magnetic Reconnection Onset via Disruption of a Forming Current Sheet by the Tearing Instability. \emph{Phys. Rev. Lett.} {\bf 116}, 105003 (2016).

    \bibitem{Samtaney2009}
        Samtaney, R. \emph{et al.} Formation of plasmoid chains in magnetic reconnection. \emph{Phys. Rev. Lett.} {\bf 103}, 105004 (2009).

    \bibitem{Uzdensky2010}
        Uzdensky, D. A., Loureiro, N. F. \& Schekochihin, A. A. Fast magnetic reconnection in the plasmoid-dominated regime. \emph{Phys. Rev. Lett.} {\bf 105}, 235002 (2010).

    \bibitem{Dong2012}
        Dong, Q.-L. \emph{et al.} Plasmoid Ejection and Secondary Current Sheet Generation from Magnetic Reconnection in Laser-Plasma Interaction. \emph{Phys. Rev. Lett.} {\bf 108}, 215001 (2012).

    \bibitem{Hare2017}
        Hare, J. D. \emph{et al.} Anomalous Heating and Plasmoid Formation in a Driven Magnetic Reconnection Experiment. \emph{Phys. Rev. Lett.} {\bf118}, 085001 (2017).

    \bibitem{Petrasso2009}
        Petrasso, R. D. \emph{et al.} Lorentz mapping of magnetic fields in hot dense plasmas. \emph{Phys. Rev. Lett.} {\bf 103}, 085001 (2009).

    \bibitem{Gao2015}
        Gao, L. \emph{et al.} Precision mapping of laser-driven magnetic fields and their evolution in high-energy-density plasmas. \emph{Phys. Rev. Lett.} {\bf 114}, 215003 (2015).

    \bibitem{Zhao2021}
        Zhao, Z. H. \emph{et al.} Three-dimensional synchronous proton radiography for dynamic magnetic fields in laser-produced high-energy-density plasmas. {\bf submitted}.

    \bibitem{Bott2017}
        Bott, A. F. A. \emph{et al.} Proton imaging of stochastic magnetic fields. \emph{J. Plasma Phys.} {\bf 83}, 905830614 (2017).

    \bibitem{Graziani2017}
        Graziani, C., Tzeferacos, P., Lamb, D. Q. \& Li, C. Inferring morphology and strength of magnetic fields from proton radiographs. \emph{Rev. Sci. Instrum}. {\bf 88}, 123507 (2017).

    \bibitem{Tubman2021}
        Tubman, E. R. \emph{et al}. Observations of pressure anisotropy effects within semi-collisional magnetized plasma bubbles. \emph{Nat. Commun.} {\bf 12}, 334 (2021).

    \bibitem{Fryxell2000}
        Fryxell, B. \emph{et al.} FLASH : AN ADAPTIVE MESH HYDRODYNAMICS CODE FOR MODELING ASTROPHYSICAL THERMONUCLEAR FLASHES. \emph{The Astrophysical Journal Supplement Series} {\bf 131}, 273-334 (2000).

    \bibitem{Heltemes2012}
        Heltemes, T. A. \& Moses, G. A. BADGER v1.0: A Fortran equation of state library. \emph{Comput. Phys. Commun.} {\bf 183}, 2629–2646 (2012).

    \bibitem{Macfarlane1989}
        Macfarlane, J. J. IONMIX-A CODE FOR COMPUTING THE EQUATION OF STATE AND RADIATIVE PROPERTIES OF LTE AND NON-LTE PLASMAS. \emph{Comput. Phys. Commun.} {\bf 56}, 259–278 (1989).
    \bibitem{Fox2011}
        Fox, W., Bhattacharjee, A. \& Germaschewski, K. Fast Magnetic Reconnection in Laser-Produced Plasma Bubbles. \emph{Phys. Rev. Lett.} {\bf 106}, 215003 (2011).

    \bibitem{Totorica2016}
        Totorica, S. R., Abel, T. \& Fiuza, F. Nonthermal Electron Energization from Magnetic Reconnection in Laser-Driven Plasmas. \emph{Phys. Rev. Lett.} {\bf 116}, 095003 (2016).

    \bibitem{Xu2016}
        Xu, Z. \emph{et al.} Characterization of magnetic reconnection in the high-energy-density regime. \emph{Phys. Rev. E} {\bf 93}, 033206 (2016).

    \bibitem{Remington1999}
        Remington, B. A., Arnett, D., Drake, R. P. \& Takabe, H. Modeling Astrophysical Phenomena in the Laboratory with Intense Lasers. \emph{Science} {\bf 284}, 1488–1493 (1999).

    \bibitem{Ryutov2000}
        Ryutov, D. D., Drake, R. P. \& Remington, B. A. Criteria for Scaled Laboratory Simulations of Astrophysical MHD Phenomena. \emph{Astrophys. J. Suppl. Ser.} {\bf 127}, 465–468 (2000).

    \bibitem{Ryutov2001}
        Ryutov, D. D., Remington, B. A., Robey, H. F., Drake, R. P. \& Introduction, I. Magnetohydrodynamic scaling: From astrophysics to the laboratory*. \emph{Phys. Plasmas} {\bf 8}, 1804 (2001).

    \bibitem{Nanbu1998}
        Nanbu, K. \& Yonemura, S. Weighted Particles in Coulomb Collision Simulations Based on the Theory of a Cumulative Scattering Angle. \emph{J. Comput. Phys.} {\bf 145}, 639-654 (1998).

    \bibitem{Arber2015}
        Arber, T. D. \emph{et al.} Contemporary particle-in-cell approach to laser-plasma modelling. \emph{Plasma. Phys. Controlled. Fusion.} {\bf 57}, 113001  (2015).

    \bibitem{Fitzpatrick2004}
        Fitzpatrick, R. Collisionless magnetic reconnection with arbitrary guide field. \emph{Phys. Plasmas} {\bf 11}, 4713 (2004).

    \bibitem{Loureiro2017}
        Loureiro, N. F. \& Boldyrev, S. Collisionless Reconnection in Magnetohydrodynamic and Kinetic Turbulence. \emph{Astrophys. J.} {\bf 850}, 182 (2017).

    \bibitem{Hosseinpour2014}
        Hosseinpour, M. A fast tearing mode instability driven by agyrotropic electron pressure. \emph{Adv. Sp. Res.} {\bf 54}, 955–960 (2014).

    \bibitem{NIST}
        NIST Standard Reference Database, \url{https://www.nist.gov/pml/stopping-power-range-tables-electrons-protons-and-helium-ions}, (2017).

\end{thebibliography}
\end{document}


\title{SUPPLEMENTARY INFORMATION \\
\smallskip\smallskip\smallskip
Laboratory observation of plasmoid-dominated magnetic reconnection in hybrid collisional-collisionless regime \smallskip}

\author{Z.~H.~Zhao$^{\dagger}$}
\affiliation{Center for Applied Physics and Technology, HEDPS, and SKLNPT, School of Physics, Peking University, Beijing 100871, China}
\author{H.~H.~An$^{\dagger}$}
\affiliation{Shanghai Institute of Laser Plasma, CAEP, Shanghai 201800, China}
\author{Y.~Xie}
\affiliation{Center for Applied Physics and Technology, HEDPS, and SKLNPT, School of Physics, Peking University, Beijing 100871, China}
\author{Z.~Lei}
\affiliation{Center for Applied Physics and Technology, HEDPS, and SKLNPT, School of Physics, Peking University, Beijing 100871, China}
\author{W.~P.~Yao}
\affiliation{Center for Applied Physics and Technology, HEDPS, and SKLNPT, School of Physics, Peking University, Beijing 100871, China}
\author{W.~Q.~Yuan}
\affiliation{Center for Applied Physics and Technology, HEDPS, and SKLNPT, School of Physics, Peking University, Beijing 100871, China}
\author{J.~Xiong}
\affiliation{Shanghai Institute of Laser Plasma, CAEP, Shanghai 201800, China}
\author{C.~Wang}
\affiliation{Shanghai Institute of Laser Plasma, CAEP, Shanghai 201800, China}
\author{J.~J.~Ye}
\affiliation{Shanghai Institute of Laser Plasma, CAEP, Shanghai 201800, China}
\author{Z.~Y.~Xie}
\affiliation{Shanghai Institute of Laser Plasma, CAEP, Shanghai 201800, China}
\author{Z.~H.~Fang}
\affiliation{Shanghai Institute of Laser Plasma, CAEP, Shanghai 201800, China}
\author{A.~L.~Lei}
\affiliation{Shanghai Institute of Laser Plasma, CAEP, Shanghai 201800, China}
\author{W.~B.~Pei}
\affiliation{Shanghai Institute of Laser Plasma, CAEP, Shanghai 201800, China}
\author{W.~M.~Zhou$^{\textrm{\Letter}}$}
\affiliation{Science and Technology on Plasma Physics Laboratory, Research Center of Laser Fusion, China Academy of Engineering Physics (CAEP), Mianyang 621900, China}
\author{W.~Wang$^{\textrm{\Letter}}$}
\affiliation{Shanghai Institute of Laser Plasma, CAEP, Shanghai 201800, China}
\author{S.~P.~Zhu$^{\textrm{\Letter}}$}
\affiliation{Institute of Applied Physics and Computational Mathematics, Beijing 100094, China}
\author{B. Qiao$^{\textrm{\Letter}}$}
\affiliation{Center for Applied Physics and Technology, HEDPS, and SKLNPT, School of Physics, Peking University, Beijing 100871, China}
\renewcommand{\thefootnote}{}
\footnotetext{$\dagger$ \  These authors have contributed to this work equally.}
\footnotetext{$\textrm{\Letter}$ \  Corresponding authors: \url{bqiao@pku.edu.cn},  \url{zhu\_shaoping@iapcm.ac.cn}, \url{wei_wang@fudan.edu.cn}, \url{zhouwm@caep.cn}.}

\date{\today}

\begin{abstract}
\end{abstract}

\maketitle

\section*{Static proton radiographys}
\begin{figure}[htp]
    \includegraphics[width= 15.6 cm]{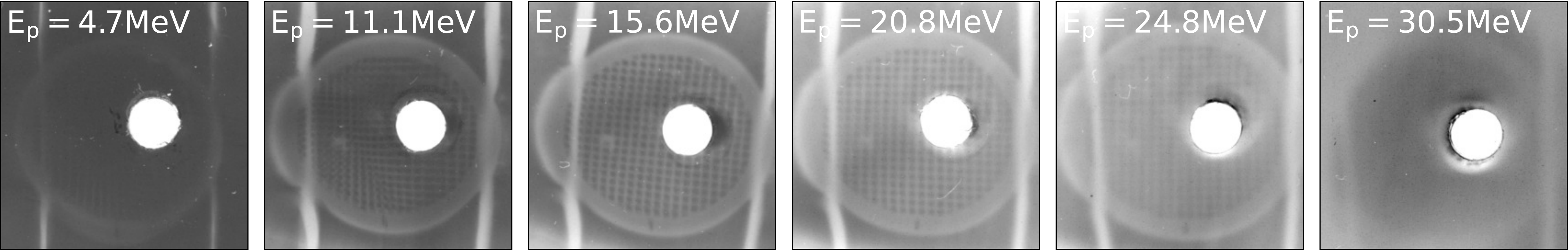}
    \caption{Static proton radiography for mesh-grids.}
    \label{suppfig:1}
\end{figure}

The high-spatial resolution of proton radiography originates from the uniform laminar proton beam produced by target normal sheath acceleration (TNSA) with relativistic ps laser pulse. Figure \ref{suppfig:1} shows the static proton radiography image for the 200-mesh-grids only in our experiment. It can be seen that the mesh grids on all radiographs are clear and distinguishable for up to 6 RCFs, which implies that the proton beams are almost uniform with cut-off energies more than 30 MeV. Therefore, such high-quality uniform proton beam is capable of dynamic radiography. It should be noted that the excessive blur comes from the scattering effect caused by the Al filters between the RCFs. In dynamic radiography shots, these filters are removed.

\section*{Three-dimensional RMHD simulation results}
\begin{figure}[htp]
    \includegraphics[width= 10 cm]{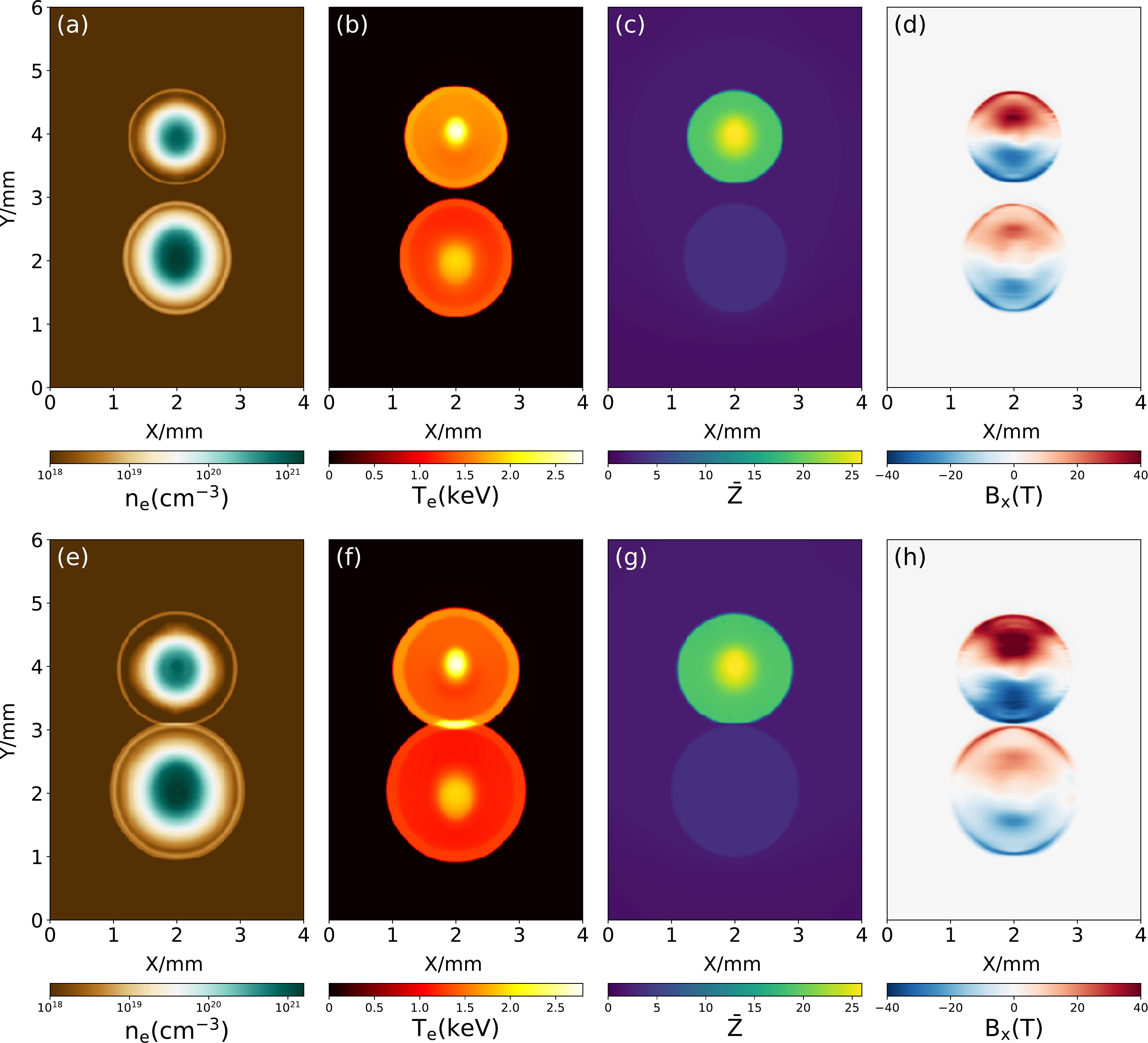}
    \caption{{\bf Three-dimensional RMHD simulation results of the laser-ablated Cu and CH plasma states}. Distributions in x-y plane of electron density $n_e$, electron temperature $T_e$, ion charge state $\bar{Z}$ and self-generated Biermann magnetic fields $B_x$ are shown from left to right at time $t = 0.8$ (upper row) and $t = 1.0 $ (lower row) ns, respectively.}
    \label{suppfig:2}
\end{figure}

The three-dimensional RMHD simulation results of hybrid magnetic reconnection are shown in Figure \ref{suppfig:2}. The differences in the equation of state (EOS) and opacity of the Cu and CH lead to the asymmetric plasma states on both sides. Comparing the expanding Cu and CH plasmas, Cu plasma has lower electron density $n_e$ and higher electron temperature $T_e$. The density decreases from $8.0 \times 10^{20}$ cm$^{-3}$ near the focal spot to $2.5 \times 10^{18} $ cm$^{-3}$ in the corona, and the temperature decreases from 2.7 keV to 1.4 keV. While in CH plasma, the density decreases from $1.6 \times 10^{21}$ cm$^{-3}$ near the focal spot to $8.0 \times 10^{18} $ cm$^{-3}$ in the corona, and the temperature decreases from 1.9 keV to 1.1 keV. Under such a high temperature, the average charge state $\bar{Z}$ of Cu decreases from 26 near the center to 20 in the outside, and the CH plasma is almost completely ionized, $\bar{Z} \sim 3.5$. The steeper temperature gradient in Cu plasma produces a stronger self-generated Biermann magnetic fields, about 40 T, while that in CH plasma is about 25 T. The strong magnetic fields surround the outside of the plasma bubbles, which is consistent with the phenomenon in the experiments. The average plasma expansion velocity is also in the same order of magnitude as the experimental estimate, 800 km/s in Cu plasma and 1000 km/s in CH plasma. And at $ t = 1.0 $ ns,  the Cu and CH plasmas begin to touch and squeeze each other. The subsequent magnetic reconnection process has exceeded the scope of RMHD simulation and will be handed over to kinetic PIC simulations. 

\section*{Plasma parameters of 2D3V kinetic PIC simulations for hybrid magnetic reconnection}

\begin{figure}[htp]
    \includegraphics[width= 9.6 cm]{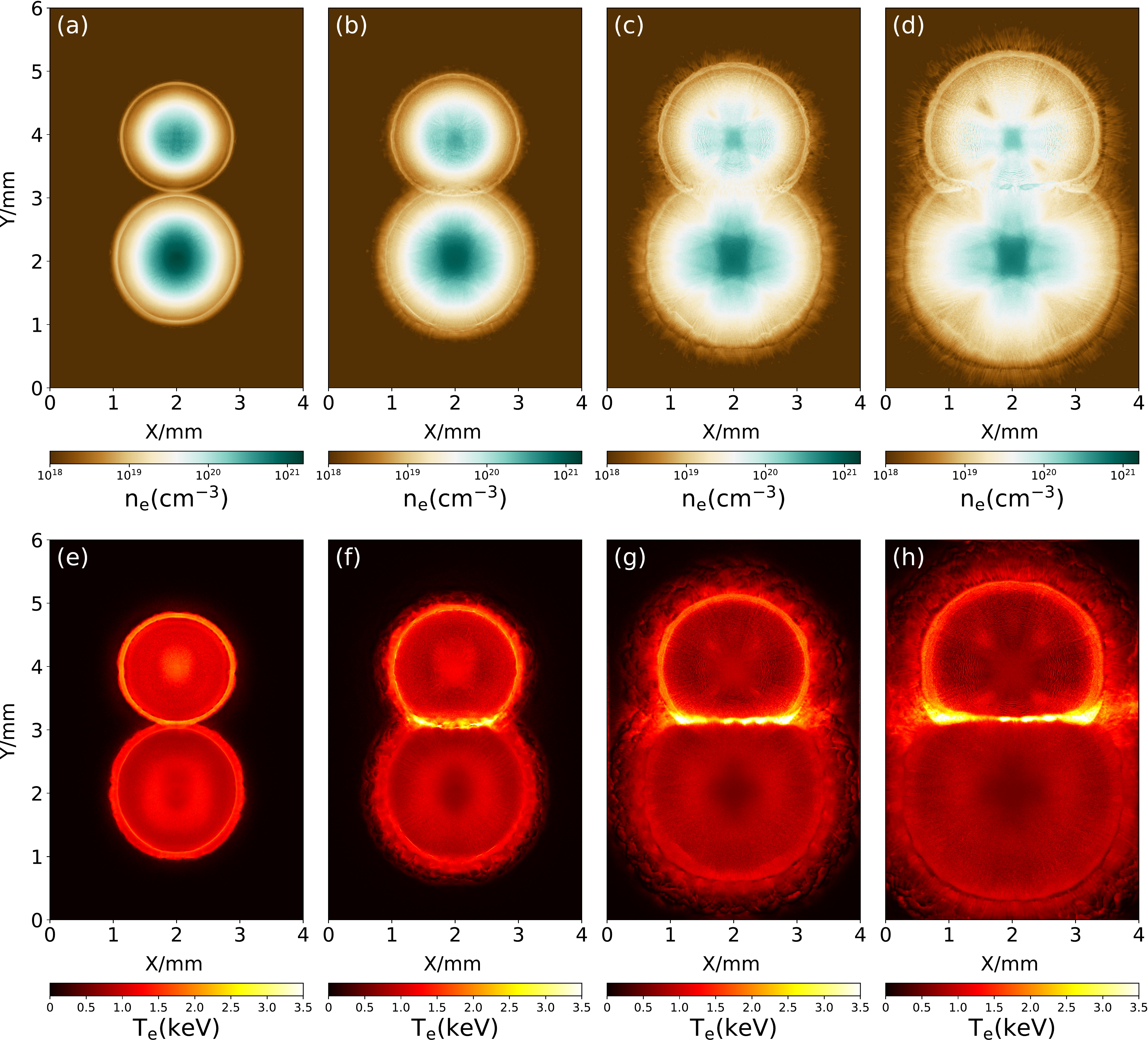}
    \caption{{\bf 2D3V PIC simulation results of hybrid magnetic reconnection.} The first row and the second row respectively show the snapshots of electron density $ n_e $ and electron temperature $T_e$ at $t = $ 1.0, 1.2, 1.5 and 1.8 ns.
    }
    \label{suppfig:3}
\end{figure}

Distributions of electron density $n_e$ and electron temperature $T_e$ for both Cu and CH plasmas at various times obtained from PIC simulations are shown in Figure \ref{suppfig:3}. We see that at $t = 1.0$ ns, the Cu and CH plasma bubbles start to contact each other, as seen in Figure \ref{suppfig:2}(e) and \ref{suppfig:2}(f) in the RMHD simulation. The slower expansion velocity and lower temperature of plasmas in PIC simulation, comparing with RMHD simulation, come from the lack of continuous laser ablation after $ t = 0.8 $ ns. At later time, as the upstream supersonic plasmas continue to flow in, both plasmas are piled up and strongly compressed in the reconnection layer, resulting in increases of both electron density $n_e$ and electron temperature $T_e$. It is shown that the electron density $n_e$ increases from $10^{19}$ cm$^{-3}$ to $10^{20}$ cm$^{-3}$, and the electron temperature $T_e$ increases from 1.0 keV to 3.5 keV. When calculating the plasma parameters in the reconnection layer for magnetic reconnection, the typical electron density and temperature are taken as $n_e \sim 4 \times 10^{19}$ cm$^{-3}$ and $T_e \sim 2.5 $ keV respectively.

\section*{Comparison with purely collisional and purely collisionless magnetic reconnections}

\begin{figure}[htp]
    \includegraphics[width= 9.6 cm]{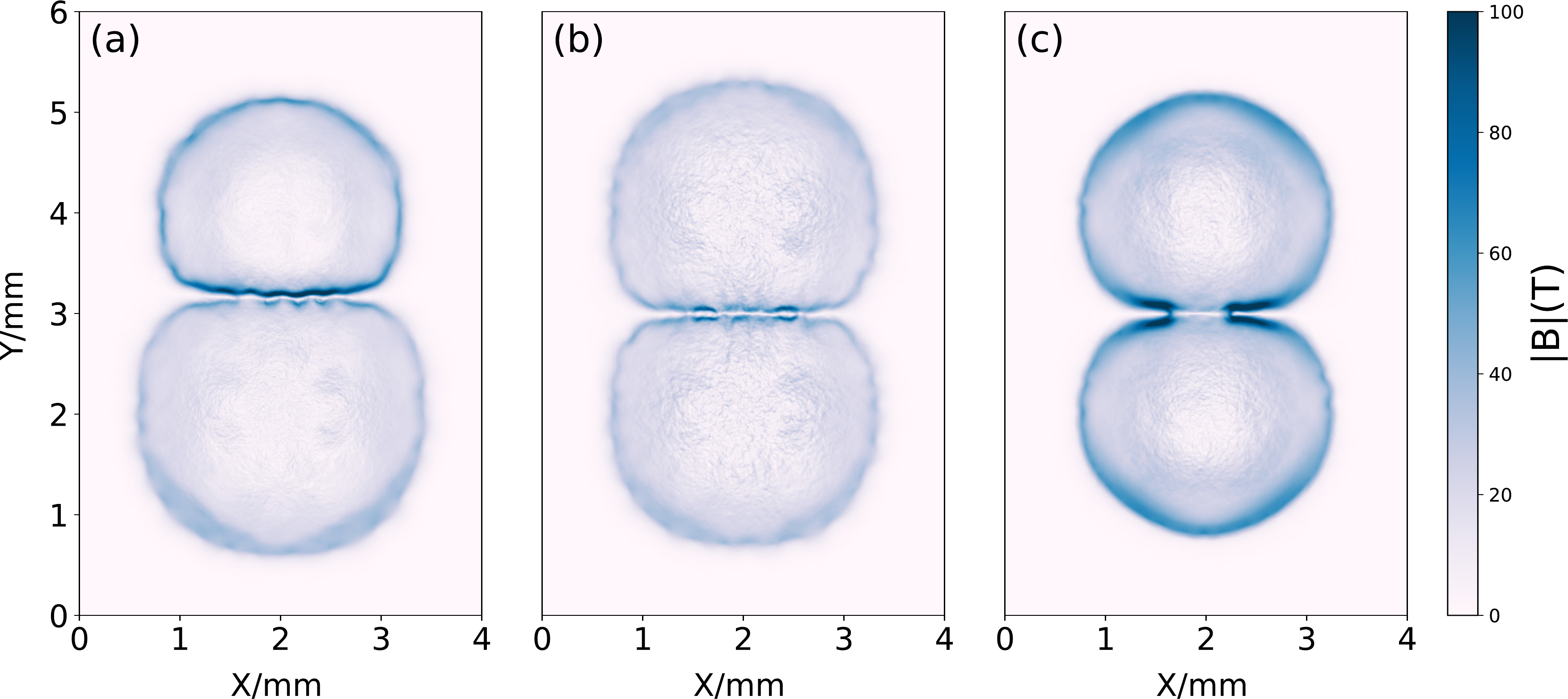}
    \caption{{\bf Comparison of hybrid collisonal-collisionless magnetic reconnection with those in purely collisional and purely collisionless regimes from PIC simulations.} The magnetic field strength $|B|$ of three different regimes of magnetic reconnections: (a) hybrid collisonal-collisionless Cu-CH reconnection at at $ t = 1.5 $ ns, (b) purely collisionless CH-CH reconnection at at $ t = 1.4 $ ns, (c) purely collisional Cu-Cu reconnection at at $ t = 1.6 $ ns. The different moments come from the time when the plasma bubbles start to contact each other.}
    \label{suppfig:4}
\end{figure}

To further demonstrate the dynamics of our hybrid reconnection, we also carry out the PIC simulations for the cases of the purely collisionless CH-CH and purely collisional Cu-Cu magnetic reconnections, where the drive laser and other parameters are also the same. The results are shown in Figures \ref{suppfig:4} for respectively (a) hybrid collisional-collisionless Cu-CH reconnection, (b) purely collisionless CH-CH reconnection and (c) purely collisional Cu-Cu reconnection.

